\documentclass[12pt,a4]{article}
\usepackage{amssymb}

\begin{document}

\begin{titlepage}
\begin{center}

\vspace*{10mm}

{\LARGE\bf
One-loop masses of open-string scalar fields in String Theory}

\vspace*{20mm}

{\large
Noriaki Kitazawa
}
\vspace{6mm}

{\it
Department of Physics, Tokyo Metropolitan University,\\
Hachioji, Tokyo 192-0397, Japan\\
e-mail: kitazawa@phys.metro-u.ac.jp
}

\vspace*{15mm}

\begin{abstract}
In phenomenological models with D-branes,
 there are in general open-string massless scalar fields,
 in addition to closed-string massless moduli fields
 corresponding to the compactification.
It is interesting to focus on the fate of such scalar fields
 in models with broken supersymmetry,
 because no symmetry forbids their masses.
The one-loop effect may give non-zero masses to them,
 and in some cases mass squared may become negative,
 which means the radiative gauge symmetry breaking.
In this article
 we investigate and propose a simple method
 for calculating the one-loop corrections
 using the boundary state formalism.
There are two categories of massless open-string scalar fields.
One consists
 the gauge potential fields corresponding to compactified directions,
 which can be understood as scalar fields in uncompactified space-time
 (related with Wilson line degrees of freedom).
The other consists ``gauge potential fields''
 corresponding to transverse directions of D-brane,
 which emerge as scalar fields in D-brane world-volume
 (related with brane moduli fields).
The D-brane boundary states
 with constant backgrounds of these scalar fields are constructed,
 and one-loop scalar masses are calculated in the closed string picture.
Explicit calculations are given in the following four concrete models:
 one D$25$-brane with a circle compactification in bosonic string theory,
 one D$9$-brane with a circle compactification in superstring theory,
 D$3$-branes at a supersymmetric ${\bf C}^3/{\bf Z}_3$ orbifold singularity,
 and a model of brane supersymmetry breaking
 with D$3$-branes and anti-D$7$-branes
 at a supersymmetric ${\bf C}^3/{\bf Z}_3$ orbifold singularity.
We show that
 the sign of the mass squared has a strong correlation with
 the sign of the related open-string one-loop vacuum amplitude.
\end{abstract}

\end{center}
\end{titlepage}

\section{Introduction}
\label{introduction}

It is important to investigate
 the fate of tree-level massless scalar fields in string models,
 because no elementary massless scalar field has been observed.
Recent developments on the moduli stabilization
 (for a review, see Ref.\cite{Douglas:2006es})
 mainly focus on closed-string massless scalar fields:
 complex structure moduli, K\"ahler moduli and dilaton.
In supersymmetric models with D-branes
 open-string massless scalar fields,
 which are superpartners of the corresponding fermion fields,
 may become massive through spontaneous supersymmetry breaking
 (for a review of string models with D-branes,
  see Ref.\cite{Blumenhagen:2006ci}).
In D-brane models with broken supersymmetry
 there are open-string massless scalar fields
 which are not the superpartners of any fermion fields.
Such scalar fields may become massive
 through the one-loop radiative correction,
 and it is possible that their squared masses become negative
 and the radiative gauge symmetry breaking is triggered.
This is a very interesting possibility
 for TeV-scale string models \cite{Antoniadis:1990ew}
 (for a review, see refs.\cite{Antoniadis:2004wm,Antoniadis:2007uz}),
 in which such a scalar field can be identified as
 the Higgs doublet field for the electroweak symmetry breaking
 \cite{Antoniadis:2000tq,Kitazawa:2006if}.
This can be a candidate of the mechanism of
 natural and necessary electroweak symmetry breaking.

Although the construction of phenomenological models with D-branes
 becomes possible by the modern understanding of open strings
 \cite{Sagnotti:1987tw,Pradisi:1988xd,Horava:1989vt,Horava:1989ga,
       Bianchi:1990yu,Bianchi:1990tb,Bianchi:1991eu,Gimon:1996rq}
 (for a review, see Refs.\cite{Dudas:2000bn,Angelantonj:2002ct}),
 the technique for actual calculations of amplitudes,
 for example, ``two-point functions'' for the mass,
 remains technically complicated in general.
For example,
 the concrete construction and usage of open string vertex operators
 are not simple straightforward tasks,
 and the integration over the places of the vertex operators
 are sometimes non-trivial requiring regularization of divergences
 or subtraction of physical divergences with some physical interpretations
 (see Ref. \cite{Green:1987mn} for the simplest case). 

In Ref.\cite{Callan:1988wz} one-loop corrections to
 equations of motion of open-string low-energy effective fields
 are investigated using the closed-string boundary state formalism.
The actual calculations are simplified
 by going to the closed string picture from the open string picture,
 because the one-loop effects in the open string picture can be understood
 as the tree-level propagations of closed strings
 between D-branes (and orientifold fixed planes)
 due to the open-closed string duality.
The main non-trivial point is
 the construction of closed-string boundary states
 which include open-string backgrounds.
In Ref.\cite{Callan:1988wz}
 the boundary state with the general gauge field background is constructed
 assuming that open strings can propagate in all the space dimensions
 (open strings with Neumann boundary condition in all the space directions,
 or space-filling D-brane).
The procedure is the following.
First, ``boundary coordinate operators'' are identified,
 and the eigenstates of those operators are constructed.
The eigenvalues are called ``boundary coordinates''.
Next, the ``boundary action''
 corresponding to the general background gauge field is identified.
The boundary action is described by boundary coordinates.
Finally, the boundary state is obtained
 by integrating the eigenstates over boundary coordinates
 with the weight of the boundary action.
It is possible
 to incorporate open-string Dirichlet boundary condition
 through simple duality arguments.
It is also possible to identify the boundary action
 corresponding to the background open-string scalar field
 through the observation that open-string massless scalar fields
 correspond to some special components of gauge potential fields.

The boundary action with the general background gauge field 
 in Ref.\cite{Callan:1988wz} (in Euclidean metric) is
\begin{equation}
 S_A^{\rm CLNY} = {i \over {4\pi}} \int_0^{2\pi} d\sigma
  \left[ A_\mu(X) \partial_\sigma X^\mu
         - {1 \over 2} i F_{\mu\nu}(X) \theta^\mu \theta^\nu
  \right]_{\tau=0},
\label{boundary-action-CLNY}
\end{equation}
 where $\sigma$ and $\tau$ are world-sheet coordinates
 ($\tau=0$ denotes the boundary of closed strings),
 $X^\mu$ are bosonic ``boundary fields''
 described by bosonic boundary coordinates,
 $\theta^\mu$ are fermionic boundary fields
 described by fermionic boundary coordinates,
 $A_\mu$ are gauge potential fields
 and $F_{\mu\nu}$ is the gauge field strength field.
It is described in Ref.\cite{Callan:1988wz} that this boundary action
 ``represents a condensate of photon vertices''. 
In case of constant gauge potential field,
 which we discuss in the next section,
 it is simple to take a part of Polyakov action
 with the background gauge field (in Minkowski metric)
\begin{equation}
 S_A = - \int_0^{2\pi} d\sigma
       \left[ A_\mu \partial_\sigma X^\mu \right]_{\tau=0}.
\label{boundary-action-Polyakov}
\end{equation}
This actually corresponding to
 the first term of Eq.(\ref{boundary-action-CLNY})
 with a different normalization.
The normalization of the gauge field in Eq.(\ref{boundary-action-Polyakov})
 is clearly known,
 because the action is used to derive the D$p$-brane effective action
 (Dirac-Born-Infeld (DBI) action)
\begin{equation}
 S_p = - T_p^{\rm DBI} \int d^{p+1}\xi \,
       e^{-\Phi}
       \sqrt{-\det \left( G + B + 2 \pi \alpha' F \right)},
\label{DBI-action}
\end{equation}
 where $T_p^{\rm DBI}$ is the tension of D$p$-brane,
 and $\Phi$, $G$ and $B$ are closed string massless fields.

At first glance
 the boundary action of Eq.(\ref{boundary-action-Polyakov})
 seems always to be zero in case of the constant gauge potential field,
 because the closed-string bosonic world-sheet fields $X^\mu$
 should have the period $2\pi$ in $\sigma$.
This actually corresponds to the fact
 that the constant gauge potential field is not physical by itself
 in ordinary situations.
There are two situations
 in which $X^\mu$ may not have the period $2\pi$ in $\sigma$.
In case of that some space dimensions are compactified,
 $X^\mu$ of compactified directions do not necessarily
 have the period $2\pi$ in $\sigma$,
 because of the existence of closed string winding modes.
In this case the corresponding components of the gauge potential field
 can be identified to the scalar fields
 which are related with Wilson line degrees of freedom
 \cite{Antoniadis:2000tq}.
In case that a stack of D-branes is at some orbifold singularity,
 $X^\mu$ of directions perpendicular to the D-branes
 do not necessarily have the period $2\pi$ in $\sigma$,
 because of the existence of twisted closed strings.
In this case the corresponding components of the gauge potential field
 also can be identified to the scalar fields
 which are related with brane moduli fields
 \cite{Kitazawa:2006if}.

In the next section,
 we review the construction of boundary states
 by the method of Ref.\cite{Callan:1988wz},
 and apply the method to construct the boundary states with
 the open-string background fields corresponding to the above two cases.
We consider three concrete models for simple explanations of the method
 to calculate the one-loop mass of the scalar field.
In section \ref{winding}
 for the case of winding closed string exchanges,
 we consider two models:
 one D$25$-brane with a circle compactification in bosonic string theory
 and
 one D$9$-brane with a circle compactification in type IIB superstring theory.
In section \ref{twisted}
 for the case of twisted closed string exchanges,
 the model of D$3$-branes at a supersymmetric ${\bf C}^3/{\bf Z}_3$
 orbifold fixed point (or singularity) is considered.
The one-loop corrections to the scalar masses are concretely calculated
 in each section.
In section \ref{brane-susy-br}
 we apply the method to a simple but non-trivial non-tachyonic model
 without supersymmetry:
 D$3$-branes and anti-D$7$-branes
 at a supersymmetric ${\bf C}^3/{\bf Z}_3$ orbifold singularity.
We encounter the divergence
 due to the tadpole couplings of massless twisted NS-NS fields with D-branes.
The application of tadpole resummations to obtain finite result
 is sketched.
In section \ref{conclusions}
 we make some comments about the application to phenomenological models.
All through the paper,
 we will be explicit in fundamental definitions
 towards future applications.

\section{Boundary states with open-string backgrounds}
\label{boundary-states}

We review the prescription of Ref.\cite{Callan:1988wz}
 for constructing closed-string boundary states
 in type IIB superstring theory with flat ten-dimensional space-time.

The mode expansion of closed-string world-sheet boson fields is
\begin{equation}
 X^\mu(\sigma,\tau)
 = {\hat x}^\mu + \alpha' {\hat p}^\mu \tau
 + i \sqrt{{{\alpha'} \over 2}}
   \sum_{m \ne 0} {1 \over m}
   \left\{
    \alpha^\mu_m e^{-i(\tau-\sigma)m}
    + {\tilde \alpha}^\mu_m e^{-i(\tau+\sigma)m}
   \right\},
\end{equation}
 and the quantization gives
\begin{equation}
 [ {\hat x}^\mu, {\hat p}^\nu ] = i \eta^{\mu\nu},
 \quad
 [ \alpha^\mu_m, \alpha^\nu_n ]
  = [ {\tilde \alpha}^\mu_m, {\tilde \alpha}^\nu_n]
  = m \delta_{m,-n} \eta^{\mu\nu}
\end{equation}
 with $(\alpha^\mu_m)^\dag = \alpha^\mu_{-m}$ and
 $(\tilde{\alpha}^\mu_m)^\dag = \tilde{\alpha}^\mu_{-m}$.
Consider the propagation of a closed string.
The corresponding world-sheet is a cylinder
 and we take the world-sheet coordinate so that $\tau=0$
 denotes one boundary of the cylinder.
At the boundary
\begin{equation}
 X^\mu(\sigma,\tau=0)
 = {\hat x}^\mu
 + \sum_{m = 1}^\infty {1 \over \sqrt{m}}
   \left[
    \left\{ a^\mu_m + {\tilde a}^\mu_{-m} \right\} e^{-im\sigma}
    +
    \left\{ {\tilde a}^\mu_m + a^\mu_{-m} \right\} e^{im\sigma}
   \right],
\end{equation}
 where
\begin{equation}
 a^\mu_m \equiv 
  \left\{
  \begin{array}{ll}
   + i \sqrt{{\alpha'} \over 2} {1 \over \sqrt{m}} {\tilde \alpha}^\mu_m
    \quad m > 0
   \\
   - i \sqrt{{\alpha'} \over 2} {1 \over \sqrt{|m|}} {\tilde \alpha}^\mu_m
    \quad m < 0
  \end{array}
  \right.,
\quad
 {\tilde a}^\mu_m \equiv 
  \left\{
  \begin{array}{ll}
   + i \sqrt{{\alpha'} \over 2} {1 \over \sqrt{m}} \alpha^\mu_m
    \quad m > 0
   \\
   - i \sqrt{{\alpha'} \over 2} {1 \over \sqrt{|m|}} \alpha^\mu_m
    \quad m < 0
  \end{array}
  \right.
\end{equation}
 with the commutation relations normalized as that for harmonic oscillators:
\begin{equation}
 [ a^\mu_m, a^\nu_{-n} ] = \delta_{m,n} \eta^{\mu\nu},
\quad
 [ {\tilde a}^\mu_m, {\tilde a}^\nu_{-n} ] = \delta_{m,n} \eta^{\mu\nu}
\end{equation}
 in the unit of $\alpha'=2$.
(We always explicitly write the $\alpha'$ dependence in this article,
 except for the above equations.)
We see that specific combinations of
\begin{equation}
 {\hat x}^\mu_m \equiv a^\mu_m + {\tilde a}^\mu_{-m},
\quad
 {\hat {\bar x}}^\mu_m \equiv {\tilde a}^\mu_m + a^\mu_{-m}
\end{equation}
 appear at the boundary,
 and these operators are called ``boundary coordinate operators''
 with the zero mode operator ${\hat x}^\mu$.
Note that the condition for the closed string
\begin{equation}
 \partial_\sigma X^\mu(\sigma,\tau=0)
 = i \sum_{m=1}^\infty \sqrt{m}
   \left\{
    - {\hat x}^\mu_m e^{-im\sigma} + {\hat {\bar x}}^\mu_m e^{im\sigma}
   \right\}
 = 0,
\end{equation}
 which corresponds to the Neumann boundary condition of the open string,
 is simply described as ${\hat x}^\mu_m={\hat {\bar x}}^\mu_m=0$.

The eigenstates of these boundary operators
 with eigenvalues $x^\mu_m$ and ${\bar x}^\mu_m$, namely
\begin{equation}
 {\hat x}^\mu_m \vert x,{\bar x} \rangle
  = x^\mu_m \vert x,{\bar x} \rangle,
\quad
 {\hat {\bar x}}^\mu_m \vert x,{\bar x} \rangle
  = {\bar x}^\mu_m \vert x,{\bar x} \rangle
\end{equation}
 for $m>0$, can be explicitly constructed as
\begin{equation}
 \vert x,{\bar x} \rangle
 = \exp\left\{-{1 \over 2}({\bar x}|x)
              -(a^\dag|{\tilde a}^\dag)
              +(a^\dag|x)
              +({\bar x}|{\tilde a}^\dag)
       \right\}
   \vert 0 \rangle, 
\label{eigenstate-X}
\end{equation}
 where
\begin{equation}
 ({\bar x}|x)
  \equiv {2 \over {\alpha'}}
         \sum_{m=1}^\infty \eta_{\mu\nu} {\bar x}^\mu_m x^\nu_m,
\end{equation}
 and the same for the others.
These states satisfy the completeness condition
\begin{equation}
 \int {\cal D}{\bar x}{\cal D}x 
      \vert x,{\bar x} \rangle \langle x,{\bar x} \vert
      = 1,
\end{equation}
 where
\begin{equation}
 {\cal D}{\bar x}{\cal D}x
  \equiv {\cal N}_x
  \prod_{\mu=0}^9 \prod_{m=1}^\infty d{\bar x}^\mu_m dx^\mu_m.
\end{equation}
We define the normalization factor ${\cal N}_x$ so that
\begin{equation}
  \int {\cal D}{\bar x}{\cal D}x e^{-({\bar x}|x)} = 1.
\end{equation}
It is easy to understand that
\begin{equation}
 \vert B_9^X \rangle = \int {\cal D}{\bar x}{\cal D}x \vert x,{\bar x} \rangle
\end{equation}
 is a part of the boundary state of space-filling D9-brane
 which satisfies
\begin{equation}
 {\hat x}^\mu_m \vert B_9^X \rangle
  = {\hat {\bar x}}^\mu_m \vert B_9^X \rangle = 0
\end{equation}
 for $m>0$.
The integral reduces to simple Gaussian integrals.

The boundary coordinates for world-sheet fermions
 are defined in a similar way.
The mode expansion of closed-string world-sheet fermion fields is
\begin{equation}
 \psi_+^\mu(\sigma,\tau)
  = \sum_{m=-\infty}^\infty \psi^\mu_{m+\kappa}
                            e^{-i(\tau+\sigma)(m+\kappa)},
\quad
 \psi_-^\mu(\sigma,\tau)
  = \sum_{m=-\infty}^\infty {\tilde \psi}^\mu_{m+\kappa}
                            e^{-i(\tau-\sigma)(m+\kappa)},
\end{equation}
 where $\kappa=1/2,0$ denote Neveu-Schwarz (NS)
 and Ramond (R) sectors, respectively.
The quantization gives
\begin{equation}
 \{\psi^\mu_r, \psi^\nu_s \}
 = \{{\tilde \psi}^\mu_r, {\tilde \psi}^\nu_s \}
 = \delta_{r,-s} \eta^{\mu\nu}
\end{equation}
 with $(\psi^\mu_r)^\dag=\psi^\mu_{-r}$
 and $({\tilde \psi}^\mu_r)^\dag={\tilde \psi}^\mu_{-r}$.

The condition for the closed string
 which corresponds to the Neumann boundary condition for the open string is
\begin{equation}
 \psi_+^\mu(\sigma,\tau=0) - \eta i \psi_-^\mu(\sigma,\tau=0) = 0,
\end{equation}
 where $\eta = \pm 1$ is the parameter to be related with the GSO projection.
The ``boundary field'' is defined as
\begin{equation}
 \theta^\mu(\sigma;\eta) \equiv
  \psi_+^\mu(\sigma,\tau=0) - \eta i \psi_-^\mu(\sigma,\tau=0)
\end{equation}
 with mode expansion
\begin{equation}
 \theta^\mu(\sigma;\eta)
  = \sum_{m=-\infty}^\infty 
    {\hat \theta}^\mu_{m+\kappa} e^{-i\sigma(m+\kappa)}.
\end{equation}
Substituting mode expansions to the boundary condition gives
 ($r={\bf Z}+\kappa$)
\begin{equation}
 {\hat \theta}^\mu_r = \psi^\mu_r - \eta i {\tilde \psi}^\mu_{-r} = 0.
\end{equation}
The boundary coordinate operators ${\hat \theta}^\mu_r$ are defined
 such that they vanish at the boundary by the boundary condition
 in accordance with
 that for ${\hat x}^\mu_m$ and ${\hat {\bar x}}^\mu_m$.
We may drop the $\eta$ dependence of ${\hat \theta}^\mu_r$,
 since the algebra
\begin{equation}
 \{ {\hat \theta}^\mu_r, {\hat \theta}^\nu_s \} = 0,
\quad
 \{ {\hat {\bar \theta}}^\mu_r, {\hat \theta}^\nu_s \} = 0
\end{equation}
 has no dependence on $\eta$,
 where $r,s > 0$
 and ${\hat {\bar \theta}}^\mu_r \equiv {\hat \theta}^\mu_{-r}$.
There are
 zero mode boundary coordinate operators
 ${\hat \theta}^\mu_0$ in Ramond sector
 whose multiplications to the Ramond ground state give spinor states
 multiplied by corresponding anti-symmetric products of gamma matrices
 (see Ref.\cite{Callan:1988wz} for detail).
We will not discuss about this operators,
 because they are irrelevant to the main topics of this paper.

The eigenstates for the boundary coordinate operators
 ${\hat \theta}^\mu_r$ and ${\hat {\bar \theta}}^\mu_r$ ($r>0$)
 can be obtained as
\begin{equation}
 \vert \theta, {\bar \theta} ; \eta \rangle
 = \exp\left\{-{1 \over 2}({\bar \theta}|\theta)
              - \eta i (\psi^\dag|{\tilde \psi}^\dag)
              + (\psi^\dag|\theta)
              + \eta i ({\bar \theta}|{\tilde \psi}^\dag)
       \right\}
        \vert 0 \rangle, 
\label{eigenstate-psi}
\end{equation}
 where
\begin{equation}
 ({\bar \theta}|\theta)
  \equiv \sum_{r>0}^\infty \eta_{\mu\nu} {\bar \theta}^\mu_r \theta^\nu_r,
\end{equation}
 and the same for the others.
These states satisfy the completeness condition
\begin{equation}
 \int {\cal D}{\bar \theta}{\cal D}\theta 
      \vert \theta,{\bar \theta} ; \eta
      \rangle \langle \theta,{\bar \theta}; \eta \vert
      = 1,
\end{equation}
 where
\begin{equation}
 {\cal D}{\bar \theta}{\cal D}\theta
  \equiv {\cal N}_\theta
  \prod_{\mu=0}^9 \prod_r^\infty d{\bar \theta}^\mu_r d\theta^\mu_r.
\end{equation}
We define the normalization factor ${\cal N}_\theta$ so that
\begin{equation}
  \int {\cal D}{\bar \theta}{\cal D}\theta e^{-({\bar \theta}|\theta)} = 1.
\end{equation}
It is easy to understand that
\begin{equation}
 \vert B_9^\psi ; \eta \rangle
  = \int {\cal D}{\bar \theta}{\cal D}\theta
    \vert \theta,{\bar \theta} ; \eta \rangle
\end{equation}
 is a part of the boundary state of space-filling D9-brane
 which satisfies
\begin{equation}
 {\hat \theta}^\mu_r \vert B_9^\psi \rangle
  = {\hat {\bar \theta}}^\mu_r \vert B_9^\psi \rangle = 0
\end{equation}
 for $r>0$.
The integral, again, reduces to simple Gaussian integrals.

The contributions of world-sheet ghost fields to the boundary state
 are determined by imposing BRST invariance.
They are obtained as follows
 (see Ref.\cite{DiVecchia:1999rh} for details).
The $bc$-ghost contribution is
\begin{equation}
 \vert B_{\rm gh} \rangle
 = \prod_{n=1}^\infty
   e^{
   \left( c_{-n} {\tilde b}_{-n} - b_{-n} {\tilde c}_{-n} \right)
   }
   \cdot
   {{c_0 + {\tilde c}_0} \over 2}
   \vert 0 \rangle_{\rm bc} {\widetilde {\vert 0 \rangle}}_{\rm bc},
\end{equation}
 where the mode expansions are
\begin{equation}
 b_+(\sigma,\tau)
 = \sum_{m=-\infty}^\infty {\tilde b}_m e^{-i(\tau+\sigma)m},
\quad
 b_-(\sigma,\tau)
 = \sum_{m=-\infty}^\infty b_m e^{-i(\tau-\sigma)m},
\end{equation}
\begin{equation}
 c_+(\sigma,\tau)
 = \sum_{m=-\infty}^\infty {\tilde c}_m e^{-i(\tau+\sigma)m},
\quad
 c_-(\sigma,\tau)
 = \sum_{m=-\infty}^\infty c_m e^{-i(\tau-\sigma)m}
\end{equation}
 satisfying the algebra
\begin{equation}
 \{ b_m, c_n \} = \{ {\tilde b}_m, {\tilde c}_n \} = \delta_{m,-n},
\end{equation}
 with $(b_m)^\dag = b_{-m}$, $(c_m)^\dag = c_{-m}$, and so on,
 and the vacuum state are defined as
\begin{equation}
 b_m \vert 0 \rangle_{\rm bc} = c_n \vert 0 \rangle_{\rm bc} = 0,
\quad
 {\tilde b}_m {\widetilde {\vert 0 \rangle}}_{\rm bc}
  = {\tilde c}_n {\widetilde {\vert 0 \rangle}}_{\rm bc} = 0
\end{equation}
 for $m \ge 0$ and $n \ge 1$.
The $\beta\gamma$-ghost contributions are
\begin{equation}
 \vert B_{\rm sgh} ; \eta \rangle_{\rm NS}
 = \prod_{r=1/2}^\infty
   e^{i\eta
       \left(
        \gamma_{-r} {\tilde \beta}_{-r} - \beta_{-r} {\tilde \gamma}_{-r}
       \right)
      }
      \vert 0 \rangle_{\rm NS}^{\beta\gamma}
\end{equation}
 for Neveu-Schwarz sector and
\begin{equation}
 \vert B_{\rm sgh} ; \eta \rangle_{\rm R}
 = \prod_{r=1}^\infty
   e^{i\eta
       \left(
        \gamma_{-r} {\tilde \beta}_{-r} - \beta_{-r} {\tilde \gamma}_{-r}
       \right)
      }
      \cdot
      e^{i\eta \gamma_0 {\tilde \beta}_0}
      \vert 0 \rangle_{\rm R}^{\beta\gamma}
\end{equation}
 for Ramond sector, where the mode expansions are
\begin{equation}
 \beta_+(\sigma,\tau)
 = \sum_{m=-\infty}^\infty \beta_{m+\kappa}
    e^{-i(\tau+\sigma)(m+\kappa)},
\quad
 \beta_-(\sigma,\tau)
 = \sum_{m=-\infty}^\infty {\tilde \beta}_{m+\kappa}
    e^{-i(\tau-\sigma)(m+\kappa)},
\end{equation}
\begin{equation}
 \gamma_+(\sigma,\tau)
 = \sum_{m=-\infty}^\infty \gamma_{m+\kappa}
    e^{-i(\tau+\sigma)(m+\kappa)},
\quad
 \gamma_-(\sigma,\tau)
 = \sum_{m=-\infty}^\infty {\tilde \gamma}_{m+\kappa}
    e^{-i(\tau-\sigma)(m+\kappa)}
\end{equation}
 satisfying the algebra
\begin{equation}
 [ \gamma_r, \beta_s ] = [ {\tilde \gamma}_r, {\tilde \beta}_s ]
  = \delta_{r,-s},
\end{equation}
 with $(\beta_r)^\dag = - \beta_{-r}$, $(\gamma_r)^\dag = \gamma_{-r}$
 and so on,
 and the vacuum states are defined as
\begin{equation}
 \vert 0 \rangle_{\rm NS}^{\beta\gamma}
  = \vert P=-1 \rangle \vert {\tilde P}=-1 \rangle,
\quad
\vert 0 \rangle_{\rm R}^{\beta\gamma}
  = \vert P=-1/2 \rangle \vert {\tilde P}=-3/2 \rangle.
\end{equation}
Here, $P$ and ${\tilde P}$ denote pictures.
The adjoint states of these ghost contributions
 can be obtained by taking Hermite conjugate,
 except for the vacuum state for Ramond-sector $\beta\gamma$-ghost
 should be taken as
\begin{equation}
 {}_{\rm R}^{\beta\gamma} \langle 0 \vert
  = \langle {\tilde P}=-1/2 \vert \langle P=-3/2 \vert.
\end{equation}
The insertion of the $bc$-ghost zero mode operator
 $({\tilde c}_0 - c_0)({\tilde b}_0+b_0)$ is also understood
 in the definition of the inner product of the $bc$-ghost boundary state
 \cite{Callan:1987px}.

The boundary state for D$p$-brane for any $p$ can be obtained
 by taking the dual transformation in open-string Dirichlet directions.
Consider that $i$ is the space index
 which denotes open-string Dirichlet directions.
The boundary condition for the closed string
 corresponding to the open-string Dirichlet boundary condition
 can be obtained by the replacement of
 ${\tilde \alpha^i}_m \rightarrow - {\tilde \alpha}^i_m$
 and ${\tilde \psi^i}_r \rightarrow - {\tilde \psi}^i_r$
 for $m,r \ne 0$.
The previous procedure receives some simple replacements only.
The boundary coordinate operators
 ${\hat x}^i_m$, ${\hat {\bar x}}^i_m$,
 ${\hat \theta}^i_r$ and ${\hat {\bar \theta}}^i_r$ ($m,r > 0$)
 should be replaced by
 the corresponding ``dual boundary coordinate operators'' defined as
\begin{equation}
 {\hat x}_D{}^i_m \equiv - a^i_m + {\tilde a}^i_{-m},
\quad
 {\hat {\bar x}}_D{}^i_m \equiv {\tilde a}^i_m - a^i_{-m},
\end{equation}
\begin{equation}
 {\hat \theta}_D{}^i_r
  \equiv \psi^i_r + \eta i {\tilde \psi}^i_{-r},
\quad
 {\hat {\bar \theta}}_D{}^i_r
  \equiv \psi^i_{-r} + \eta i {\tilde \psi}^i_r,
\end{equation}
 respectively.
The operators in the states of
 Eqs.(\ref{eigenstate-X}) and (\ref{eigenstate-psi})
 also receive the appropriate replacements.
The zero mode boundary coordinate ${\hat x}^i$
 requires a special treatment.
We have to multiply
 the operator $\delta({\hat x}^i-y^i)$ for each $i$ to the boundary state,
 which ensures that the place of the center of mass of the closed string
 should be fixed at $y^i$ at the boundary.
It is easy to check that
 the obtained boundary states satisfy closed-string boundary conditions
 corresponding to open-string Dirichlet boundary conditions.

The contribution of
 the zero-mode operators of world-sheet fermion fields in the Ramond sector
 should be considered separately (see Ref.\cite{DiVecchia:1999rh} for details).
The Ramond ground state for D$p$-brane is given by
\begin{equation}
 \vert B_p^\psi ; \eta \rangle_{\rm R}^{(0)}
 = M_{AB} \vert A \rangle {\widetilde {\vert B \rangle}},
\label{Ramond-ground}
\end{equation}
 where $\vert A \rangle$ and ${\widetilde {\vert B \rangle}}$
 are ten-dimensional spinor states with spinor indices $A$ and $B$
 and
\begin{equation}
 M = C \Gamma^0 \cdots \Gamma^p
     {{1 + i \eta \Gamma^{11}} \over {1 + i \eta}}
\end{equation}
 with ten-dimensional Dirac gamma matrices $\Gamma^\mu$ and $\Gamma^i$
 with $\mu = 0,1,\cdots,p$ and $i=p+1,p+2,\cdots,9$, 
 the charge conjugation matrix
 $C \equiv \Gamma^3 \Gamma^5 \Gamma^7 \Gamma^9 \Gamma^0$
 and the chirality matrix
 $\Gamma^{11} \equiv \Gamma^0 \Gamma^1 \cdots \Gamma^9$.
This state satisfies the boundary conditions
\begin{equation}
 {\hat \theta}^\mu_0 \vert B_p^\psi ; \eta \rangle_{\rm R}^{(0)} = 0,
\quad
 {\hat \theta}_D{}^i_0 \vert B_p^\psi ; \eta \rangle_{\rm R}^{(0)} = 0.
\end{equation}
The adjoint state is obtained by taking the Hermite conjugate
\begin{equation}
 {}_{\rm R}^{(0)}\langle B_p^\psi ; \eta \vert
 = - {\widetilde {\langle B \vert}} \langle A \vert (M^\dag)_{BA},
\end{equation}
 except for the minus sign from the Fermi statistics of spinor states,
 where
\begin{equation}
 M^\dag = {{1 - i \eta \Gamma^{11}} \over {1 - i \eta}}
          \Gamma^p \cdots \Gamma^0 C.
\end{equation}

In the prescription of Ref.\cite{Callan:1988wz}
 the closed-string boundary state with constant background
 of open-string gauge potential field on D$p$-brane is obtained
 by the integral with a weight given by the boundary action
 of Eq.(\ref{boundary-action-Polyakov}).
\begin{equation}
 \vert B_p ; \eta \rangle
 = N_p
   \int {\cal D}X{\cal D}\Theta
   \ e^{iS_A} \
   \vert x, {\bar x}, x_D, {\bar x}_D \rangle
   \vert \theta, {\bar \theta}, \theta_D, {\bar \theta}_D ; \eta \rangle
   \vert B_{\rm gh} \rangle \vert B_{\rm sgh} \rangle,
\label{boundary-with-A}
\end{equation}
 where ${\cal D}X{\cal D}\Theta \equiv
        {\cal D}{\bar x}{\cal D}x {\cal D}{\bar x}_D{\cal D}x_D
        {\cal D}{\bar \theta}{\cal D}\theta
        {\cal D}{\bar \theta}_D{\cal D}\theta_D$ and
\begin{equation}
 S_A = - A_\mu \int_0^{2\pi} d\sigma
       \left[ \partial_\sigma X^\mu \right]_{\tau=0}
\label{boundary-action-const-A}
\end{equation}
 with
\begin{equation}
 \left[ \partial_\sigma X^\mu \right]_{\tau=0}
 = i \sum_{m=1}^\infty \sqrt{m}
   \left\{
    - x^\mu_m e^{-im\sigma} + {\bar x}^\mu_m e^{im\sigma}
   \right\}.
\end{equation}
The eigenstates of the boundary coordinate operators are defined as
\begin{eqnarray}
 &&
 \vert x, {\bar x}, x_D, {\bar x}_D \rangle
  =
   \exp\left\{-{1 \over 2}({\bar x}|x)_\parallel
              -(a^\dag|{\tilde a}^\dag)_\parallel
              +(a^\dag|x)_\parallel
              +({\bar x}|{\tilde a}^\dag)_\parallel
       \right\}
\\
  && \qquad \cdot \delta^{9-p}({\hat x}-y) 
    \exp\left\{-{1 \over 2}({\bar x}_D|x_D)_\perp
              +(a^\dag|{\tilde a}^\dag)_\perp
              -(a^\dag|x_D)_\perp
              +({\bar x}_D|{\tilde a}^\dag)_\perp
       \right\}
    \vert 0 \rangle, 
\nonumber
\end{eqnarray}
 with
\begin{equation}
 ({\bar x}|x)_\parallel
  \equiv {2 \over {\alpha'}}
         \sum_{m=1}^\infty
         \sum_{\mu,\nu=0}^p \eta_{\mu\nu} {\bar x}^\mu_m x^\nu_m,
\quad
 ({\bar x}_D|x_D)_\perp
  \equiv {2 \over {\alpha'}}
         \sum_{m=1}^\infty
         \sum_{i,j=p+1}^9 \delta_{ij} {\bar x}_D{}^i_m x_D{}^j_m,
\end{equation}
 and
\begin{eqnarray}
 \vert \theta, {\bar \theta}, \theta_D, {\bar \theta}_D \rangle
  &=&
   \exp\left\{-{1 \over 2}({\bar \theta}|\theta)_\parallel
              - \eta i (\psi^\dag|{\tilde \psi}^\dag)_\parallel
              + (\psi^\dag|\theta)_\parallel
              + \eta i ({\bar \theta}|{\tilde \psi}^\dag)_\parallel
       \right\}
\\
  && \cdot
   \exp\left\{-{1 \over 2}({\bar \theta}_D|\theta_D)_\perp
              + \eta i (\psi^\dag|{\tilde \psi}^\dag)_\perp
              + (\psi^\dag|\theta_D)_\perp
              - \eta i ({\bar \theta}_D|{\tilde \psi}^\dag)_\perp
       \right\}
   \vert 0 \rangle, 
\nonumber
\end{eqnarray}
 with
\begin{equation}
 ({\bar \theta}|\theta)_\parallel
  \equiv \sum_{r>0}^\infty
         \sum_{\mu,\nu=0}^p \eta_{\mu\nu} {\bar \theta}^\mu_r \theta^\nu_r,
\quad
 ({\bar \theta}_D|\theta_D)_\perp
  \equiv \sum_{r>0}^\infty
         \sum_{i,j=p+1}^9 \delta_{ij} {\bar \theta}_D{}^i_r \theta_D{}^j_r.
\end{equation}
The normalization factor $N_p$ is determined so that
 the open-string one-loop cylinder vacuum amplitude can be obtained
 using this boundary state with vanishing background field.
\begin{equation}
 N_p \equiv {{T_p} \over 2}
  \equiv {\sqrt{\pi} \over 2} \left( 4 \pi^2 \alpha' \right)^{{3-p} \over 2}. 
\end{equation}
Note that
 the coefficient of the effective action of Eq.(\ref{DBI-action})
 is $T_p^{\rm DBI}=T_p/\kappa$,
 where $(1/\kappa^2)^{1/8}$ is the reduced Planck mass
 in ten-dimensional space-time.

We have to apply the GSO projection to the boundary state.
The operation of GSO projection operators
 $e^{i \pi F}$ and $e^{i \pi {\tilde F}}$,
 where $F$ and ${\tilde F}$ are world-sheet spinor number operators
 (including the effect of $\beta\gamma$-ghost),
 are
\begin{equation}
 e^{i \pi F} \vert B_p ; \eta \rangle_{\rm NS} 
 = - \vert B_p ; -\eta \rangle_{\rm NS},
\quad
 e^{i \pi {\tilde F}} \vert B_p ; \eta \rangle_{\rm NS} 
 = - \vert B_p ; -\eta \rangle_{\rm NS},
\end{equation}
\begin{equation}
 e^{i \pi F} \vert B_p ; \eta \rangle_{\rm R} 
 = (-1)^p \vert B_p ; -\eta \rangle_{\rm R},
\quad
 e^{i \pi {\tilde F}} \vert B_p ; \eta \rangle_{\rm R} 
 = - \vert B_p ; -\eta \rangle_{\rm R}.
\end{equation}
The factor $(-1)^p$ in R-sector comes
 due to the matrix $M$ in the boundary state.
(Remember that the operation of $e^{i \pi F}$ is essentially
 equivalent to multiply the chirality matrix $\Gamma^{11}$.)
To extract the spectrum of type IIB theory
 the GSO projection should be defined as follows \cite{DiVecchia:1999rh}.
\begin{equation}
 {{1+{e^{i \pi F}}} \over 2}{{1+{e^{i \pi {\tilde F}}}} \over 2}
 \vert B_p ; +1 \rangle_{\rm NS}
 = {1 \over 2}
   \left(
    \vert B_p ; +1 \rangle_{\rm NS}
    - \vert B_p ; -1 \rangle_{\rm NS}
   \right)
   \equiv \vert B_p \rangle_{\rm NS},
\label{GSO-NS}
\end{equation}
\begin{equation}
 {{1+(-1)^p{e^{i \pi F}}} \over 2}{{1+(-1){e^{i \pi {\tilde F}}}} \over 2}
 \vert B_p ; +1 \rangle_{\rm R}
 = {1 \over 2}
   \left(
    \vert B_p ; +1 \rangle_{\rm R}
    + \vert B_p ; -1 \rangle_{\rm R}
   \right)
   \equiv \vert B_p \rangle_{\rm R}.
\label{GSO-R}
\end{equation}

For the closed-string boundary state with the constant background
 of brane moduli fields,
 the boundary action $S_A$ should be replaced by
 \cite{Polchinski:1994fq,Callan:1995xx}
\begin{equation}
S_A^\perp = - A_i \int_0^{2\pi} d\sigma
 \left[ \partial_\tau X^i \right]_{\tau=0}
\label{boundary-action-moduli}
\end{equation}
 with
\begin{equation}
 \left[ \partial_\tau X^i \right]_{\tau=0}
 = \alpha' {\hat p}^i
 + i \sum_{m=1}^\infty \sqrt{m}
   \left\{
     x_D{}^i_m e^{-im\sigma} - {\bar x}_D{}^i_m e^{im\sigma}
   \right\}.
\label{partial-tau-X}
\end{equation}
Note that
 the boundary action is now described by dual boundary coordinates.

In case of flat ten-dimensional space-time
 both $S_A$ and $S_A^\perp$ vanish,
 because $\partial_\sigma X^\mu$ and
 the oscillator part of $\partial_\tau X^i$
 have the period $2\pi$ in $\sigma$.
(The first term of Eq.(\ref{partial-tau-X})
 simply results the shift of the place of D$p$-brane:
 $y^i \rightarrow y^i - 2 \pi \alpha' A_i$.
This $A_i$ is nothing but the brane moduli field,
 and it is not that we are interested in.
See below.)
For the first case,
 if some space directions in the D$p$-brane world-volume are compactified,
 the boundary action $S_A$ does not always vanish,
 because $\partial_\sigma X^\mu$ does not necessary have
 the period $2\pi$ in $\sigma$ for closed-string winding modes.
The vector potential fields $A_\mu$
 corresponding to the compactified directions
 become scalar background fields in uncompactified D$p$-brane world-volume,
 which correspond to Wilson line degrees of freedom.
For the second case,
 if the D$p$-brane is at some orbifold singularity,
 the boundary action $S_A^\perp$ does not always vanish,
 because $\partial_\tau X^i$ do not have the period $2\pi$ in $\sigma$
 for the closed-string twisted modes.
The ``vector potential fields'', or brane moduli fields, $A_i$
 become scalar fields on the D$p$-brane.
In this case,
 the non-trivial orbifold twist action
 to open-string Chan-Paton indices are necessary
 so that $A_i$ are not completely projected out,
 and we need to consider multiple D$p$-branes.
The background fields $A_i$ should be matrix-valued in this case
 and the action weight in Eq.(\ref{boundary-with-A})
 should be replaced by ${\rm tr}({\cal P}\exp(S_A^\perp))$.
In the following two sections
 we demonstrate the actual calculations of
 the one-loop mass corrections to the scalar fields
 in these two cases in order.
The boundary state given in this section
 should be modified a little in each case.

\section{One-loop masses by winding closed string exchanges}
\label{winding}

Consider first one D$25$-brane in bosonic string theory
 with $25$th space dimension compactified in circle of radius $R$.
Since the boundary action of Eq.(\ref{boundary-action-const-A})
 only depends on the bosonic boundary coordinate,
 this is a simple and sufficient model as the first example.
The basic boundary state is that of Eq.(\ref{boundary-with-A})
 with $26$-dimensional space-time and no contributions
 from world-sheet fermions and $\beta\gamma$-ghost fields:
\begin{equation}
 \vert B_{25} \rangle
 = N_{25}
   \int {\cal D}{\bar x}{\cal D}x
   \ e^{iS_A} \
   \vert x, {\bar x} \rangle
   \vert B_{\rm gh} \rangle,
\end{equation}
 where $N_{25}$ is
\begin{equation}
 N_p = {{T_p} \over 2}
     = {1 \over 16}{\sqrt{\pi} \over 2}
       (4 \pi^2 \alpha')^{{11-p} \over 2}
\end{equation}
 with $p=25$, and
\begin{equation}
 S_A = - A_{25} \int_0^{2\pi} d\sigma
       \left[ \partial_\sigma X^{25} \right]_{\tau=0}.
\end{equation}
In the mode expansion of the world-sheet field $X^{25}$
 the integer winding number $w$ is included,
 and we consider it as a new boundary coordinate operator.
\begin{equation}
 \partial_\sigma X^{25}(\sigma,\tau=0)
 = - R {\hat w}
   + i \sum_{m=1}^\infty \sqrt{m}
   \left\{
    - {\hat x}^{25}_m e^{-im\sigma} + {\hat {\bar x}}^{25}_m e^{im\sigma}
   \right\}.
 \end{equation}
We redefine the eigenstates of the boundary coordinate operators as
\begin{equation}
 \vert x,{\bar x} ; w \rangle
 = \exp\left\{-{1 \over 2}({\bar x}|x)
              -(a^\dag|{\tilde a}^\dag)
              +(a^\dag|x)
              +({\bar x}|{\tilde a}^\dag)
       \right\}
   \vert w \rangle, 
\end{equation}
 where $\vert w \rangle$ is the eigenstate of ${\hat w}$ satisfying
\begin{equation}
 {\hat w} \vert w \rangle = w \vert w \rangle,
\quad
 \langle w' \vert w \rangle = \delta_{w'w}.
\end{equation}
The integral in the boundary action is easily performed:
\begin{equation}
 S_A = A_{25} \ 2 \pi R \ {\hat w}.
\end{equation}
Here we temporally consider the boundary action as an operator.
The boundary action with constant $A_{25}$ background
 should be obtained as follows.
\begin{eqnarray}
 \vert B_{25}; A_{25} \rangle
 &=& N_{25}
   \sum_{w=-\infty}^\infty
   \int {\cal D}{\bar x}{\cal D}x
   \ e^{i A_{25} 2 \pi R {\hat w}} \
   \vert x, {\bar x}; w \rangle
   \vert B_{\rm gh} \rangle
\nonumber\\
 &=& N_{25}
   \sum_{w=-\infty}^\infty
   e^{i A_{25} 2 \pi R w }
   \int {\cal D}{\bar x}{\cal D}x
   \vert x, {\bar x}; w \rangle
   \vert B_{\rm gh} \rangle.
\end{eqnarray}
Following the arguments in section \ref{introduction} with DBI action,
 the canonically normalized scalar field in $25$-dimensional space-time
 is defined as
 $\phi \equiv A_{25} \sqrt{2 \pi R} / g$,
 where
\begin{equation}
 {1 \over {g^2}} = T^{\rm DBI}_{25} (2 \pi \alpha')^2 {1 \over {g_s}},
\quad
 g_s \equiv e^{\langle \Phi \rangle}.
\end{equation}
Therefore, we have
\begin{equation}
 \vert B_{25}; \phi \rangle
 = N_{25}
   \sum_{w=-\infty}^\infty
   e^{i {g \over \sqrt{2 \pi R}} \phi 2 \pi R w}
   \int {\cal D}{\bar x}{\cal D}x
   \vert x, {\bar x}; w \rangle
   \vert B_{\rm gh} \rangle.
\label{boundary-state-bosonic-with-scalar}
\end{equation}
This is the closed-string boundary state
 with the constant background of an open-string scalar field.
The boundary state without the background is clearly
\begin{equation}
 \vert B_{25} \rangle
 = N_{25}
   \sum_{w=-\infty}^\infty
   \int {\cal D}{\bar x}{\cal D}x
   \vert x, {\bar x}; w \rangle
   \vert B_{\rm gh} \rangle
 \equiv \sum_{w=-\infty}^\infty \vert B_{25};w \rangle.
\end{equation}

The ``multipoint functions'' of $\phi$ are given by
\begin{equation}
 \langle B_{25}; \phi \vert D \vert B_{25}; \phi \rangle
 = \sum_{w=-\infty}^\infty
   e^{i 2 {g \over \sqrt{2 \pi R}} \phi 2 \pi R w}
   \langle B_{25}; w \vert D \vert B_{25}; w \rangle,
\end{equation}
 where
\begin{equation}
 D \equiv
   {{\alpha'} \over {4\pi}}
   \int_0^\infty dt \int_0^{2\pi} d\varphi \ z^{L_0} {\bar z}^{{\tilde L}_0} 
\end{equation}
 with $z=e^{-t}e^{i\varphi}$
 is the closed-sting propagator operator \cite{DiVecchia:1999rh}.
This is simply an weighted sum over $w$ of
 the open-string one-loop vacuum amplitudes
 corresponding to the tree-level propagations of $w$ twisted closed strings.
This is the way how
 the spectrum in the string theory appears
 in ``multipoint function'' of $\phi$ in this case.
The explicit calculations give
\begin{equation}
 \langle B_{25}; w \vert D \vert B_{25}; w \rangle
 = 2\pi^2 \cdot {{\alpha'} \over {4\pi}} \cdot N_{25}^2 (V_{25} 2 \pi R)
   \int_0^\infty ds \ e^{-{{\pi R^2} \over {2\alpha'}} w^2 s}
   {1 \over {(\eta(is))^{24}}},
\end{equation}
 where $s \equiv t/\pi$.
We obtain the ``two-point function'' as
\begin{equation}
 A_2 = - \phi^2 2 g^2 (2 \pi R)^2
  \cdot (2\pi^2 {{\alpha'} \over {4\pi}} N_{25}^2) \cdot V_{25}
  \sum_{w=-\infty}^\infty w^2
   \int_0^\infty ds \ e^{-{{\pi R^2} \over {2\alpha'}} w^2 s}
   {1 \over {(\eta(is))^{24}}},
\end{equation}
 and the mass of the scalar field in $25$-dimensional space-time
 is obtained as
\begin{equation}
 m_\phi^2 = g^2 (2 \pi R)^2
  \cdot (2\pi^2 {{\alpha'} \over {4\pi}} N_{25}^2)
  \sum_{w=-\infty}^\infty w^2
   \int_0^\infty ds \ e^{-{{\pi R^2} \over {2\alpha'}} w^2 s}
   {1 \over {(\eta(is))^{24}}}.
\end{equation}
The scale of the mass is determined by the mixture of
 the string scale $1/\sqrt{\alpha'}$
 and the scale of compactification $1/R$.
Note that only winding closed strings, $w \ne 0$, contribute.
If we could neglect the tachyon contribution,
 the mass would go to zero in the limit of $R \rightarrow 0$,
 and the scalar field would become a brane moduli field of a D$24$-brane.
In the limit of $R \rightarrow \infty$ the mass vanishes as expected,
 since the scalar field becomes a component of the gauge field
 on the D$25$-brane.
The sign of the mass squared
 has an opposite correlation with the sign of the vacuum energy
 ($-1 \times$"zero-point function" in our notation),
 which has already pointed out in Ref.\cite{Antoniadis:2000tq}.

Next, we consider in superstring theory
 one D$9$-brane with $9$th space dimension compactified
 in circle of radius $R$.
The story is completely parallel to the above bosonic string theory case,
 except for the difference of the number of space dimensions
 and inclusion of the contribution of
 world-sheet fermions and $\beta\gamma$-ghost fields.
The contribution to the two-point function
 from the closed-string NS-NS sector is
\begin{eqnarray}
 A_2^{\rm NS}
  &=& - \phi^2 2 g^2 (2 \pi R)^2
  \cdot (2 \pi^2 {{\alpha'} \over {4\pi}} N_9^2) \cdot V_9
\nonumber\\
 && \times
  \sum_{w=-\infty}^\infty w^2
  \int_0^\infty ds \ e^{-{{\pi R^2} \over {2\alpha'}} w^2 s}
  {1 \over {(\eta(is))^8}}
  \cdot {1 \over 2}
  \left\{
   \left( {{\theta_3(is)} \over {\eta(is)}} \right)^4
   - \left( {{\theta_4(is)} \over {\eta(is)}} \right)^4
  \right\}.
\end{eqnarray}
Here $1/g^2 = T^{\rm DBI}_9 (2 \pi \alpha')^2/g_s$.
The contribution from the closed-string R-R sector
 differs only the part of the theta function.
\begin{eqnarray}
 A_2^{\rm R}
  &=& - \phi^2 2 g^2 (2 \pi R)^2
  \cdot (2 \pi^2 {{\alpha'} \over {4\pi}} N_9^2) \cdot V_9
\nonumber\\
 && \times
  \sum_{w=-\infty}^\infty w^2
  \int_0^\infty ds \ e^{-{{\pi R^2} \over {2\alpha'}} w^2 s}
  {1 \over {(\eta(is))^8}}
  \cdot {1 \over 2}
  \left\{
   - \left( {{\theta_2(is)} \over {\eta(is)}} \right)^4
  \right\}.
\end{eqnarray}
The total two-point amplitude is
\begin{eqnarray}
 &A_2&
  = - \phi^2 2 g^2 (2 \pi R)^2
  \cdot (2 \pi^2 {{\alpha'} \over {4\pi}} N_9^2) \cdot V_9
\\
 && \times
  \sum_{w=-\infty}^\infty w^2
  \int_0^\infty ds \ e^{-{{\pi R^2} \over {2\alpha'}} w^2 s}
  {1 \over {(\eta(is))^8}}
  \cdot {1 \over 2}
  \left\{
   \left( {{\theta_3(is)} \over {\eta(is)}} \right)^4
   - \left( {{\theta_4(is)} \over {\eta(is)}} \right)^4
   - \left( {{\theta_2(is)} \over {\eta(is)}} \right)^4
  \right\},
\nonumber
\end{eqnarray}
 and the mass of the scalar field in $9$-dimensional space-time is
\begin{eqnarray}
 &m_\phi^2&
  = g^2 (2 \pi R)^2
  \cdot (2 \pi^2 {{\alpha'} \over {4\pi}} N_9^2)
\\
 && \times
  \sum_{w=-\infty}^\infty w^2
  \int_0^\infty ds \ e^{-{{\pi R^2} \over {2\alpha'}} w^2 s}
  {1 \over {(\eta(is))^8}}
  \cdot {1 \over 2}
  \left\{
   \left( {{\theta_3(is)} \over {\eta(is)}} \right)^4
   - \left( {{\theta_4(is)} \over {\eta(is)}} \right)^4
   - \left( {{\theta_2(is)} \over {\eta(is)}} \right)^4
  \right\}.
\nonumber
\end{eqnarray}
As expected,
 the one-loop correction to the scalar mass vanishes
 due to the supersymmetry,
 which is realized in the above result through the identity
 $\theta_3^4-\theta_4^4-\theta_2^4=0$
 as exactly the same that happens in one-loop vacuum energy.
Once the supersymmetry is broken
 the balance in the contributions from
 space-time bosons and fermions of the open string,
 or from NS-NS boson fields and R-R boson fields of the closed string,
 is also broken,
 and the non-zero value for the scalar mass emerges.
In case that supersymmetry is broken
 at the string scale $1/\sqrt{\alpha'}$,
 the scale of the mass should be determined
 by two scales: $1/\sqrt{\alpha'}$ and $1/R$.
The sign of the mass squared
 depends on how the spectrum is modified
 by the supersymmetry breaking.

\section{One-loop masses by twisted closed string exchanges}
\label{twisted}

We consider a stack of D$3$-branes
 at a supersymmetric ${\bf C}^3/{\bf Z}_3$ orbifold singularity
 as a simple example.

The ${\bf Z}_3$ transformation is defined as
\begin{equation}
 Z^{(\pm)a}(\sigma,\tau)
  \rightarrow e^{\pm 2 \pi i v_a} Z^{(\pm)a}(\sigma,\tau),
\quad
 \psi_\pm^{(\pm)a}(\sigma,\tau)
  \rightarrow e^{\pm 2 \pi i v_a} \psi_\pm^{(\pm)a}(\sigma,\tau),
\label{Z3-transformation}
\end{equation}
 where complexified world-sheet fields are defined as
\begin{equation}
 Z^{(\pm)a} \equiv {1 \over \sqrt{2}}
                   \left( X^{2a+2} \pm i X^{2a+3} \right),
\quad
 \psi_\pm^{(\pm)a} \equiv {1 \over \sqrt{2}}
                   \left( \psi_\pm^{2a+2} \pm i \psi_\pm^{2a+3} \right)
\end{equation}
 for $a=1,2,3$,
 and $v=(1/3,1/3,-2/3)$.
In ${\bf C}^3/{\bf Z}_3$ orbifold space,
 the space points which are connected by the above ${\bf Z}_3$ transformation
 are identified.
(Here, the world-sheet fields $Z^{(\pm)a}$ are identified
 to the complexified six-dimensional space coordinates.)
Therefore, there exist so called twisted closed strings
 which look like open strings with two edges
 identified by ${\bf Z}_3$ transformations.
The world-sheet fields of such twisted closed strings
 should satisfy the conditions of
\begin{eqnarray}
 Z^{(\pm)a}(\sigma+2\pi,\tau)
  &=& e^{\pm 2 \pi i v_a} Z^{(\pm)a}(\sigma,\tau),
\\
 \psi_+^{(\pm)a}(\sigma+2\pi,\tau)
  &=& e^{-2 \pi i \kappa} e^{\pm 2 \pi i v_a} \psi_+^{(\pm)a}(\sigma,\tau),
\\
 \psi_-^{(\pm)a}(\sigma+2\pi,\tau)
  &=& e^{+2 \pi i \kappa} e^{\pm 2 \pi i v_a} \psi_-^{(\pm)a}(\sigma,\tau),
\label{twist-condition-1}
\end{eqnarray}
 or
\begin{eqnarray}
 Z^{(\pm)a}(\sigma+2\pi,\tau)
  &=& (e^{\pm 2 \pi i v_a})^2 Z^{(\pm)a}(\sigma,\tau),
\\
 \psi_+^{(\pm)a}(\sigma+2\pi,\tau)
  &=& e^{-2 \pi i \kappa} (e^{\pm 2 \pi i v_a})^2 \psi_+^{(\pm)a}(\sigma,\tau),
\\
 \psi_-^{(\pm)a}(\sigma+2\pi,\tau)
  &=& e^{+2 \pi i \kappa} (e^{\pm 2 \pi i v_a})^2 \psi_-^{(\pm)a}(\sigma,\tau).
\label{twist-condition-2}
\end{eqnarray}
Namely, there are two kinds of twisted closed strings.
In the following
 we only consider the twisted closed strings
 which satisfy the first set of conditions,
 because another kind of twisted closed strings gives the same results.
Since these conditions change
 the mode expansion of world-sheet fields,
 we have to reconstruct the boundary state,
 though the prescription is the same of
 that explained in section \ref{boundary-states}.

The mode expansion of twisted world-sheet boson fields $Z^{(\pm)a=1}$,
 we simply write this fields as $Z^{(\pm)}$ for a while,
 are given as
\begin{eqnarray}
 Z^{(+)}(\sigma,\tau)
 &=& i \sqrt{{{\alpha'} \over 2}}
   \sum_{m=-\infty}^\infty
   \Bigg\{
    {1 \over {m+1/3}} \, \alpha_{m+1/3} \, e^{-i(\tau-\sigma)(m+1/3)}
\nonumber\\
 && \qquad\qquad\qquad\qquad
     + {1 \over {m-1/3}} \, {\tilde \alpha}_{m-1/3} \,
       e^{-i(\tau+\sigma)(m-1/3)}
   \Bigg\},
\\
 Z^{(-)}(\sigma,\tau)
 &=& -i \sqrt{{{\alpha'} \over 2}}
   \sum_{m=-\infty}^\infty
   \Bigg\{
    {1 \over {m+1/3}} \, \alpha_{-(m+1/3)} \, e^{i(\tau-\sigma)(m+1/3)}
\nonumber\\
 && \qquad\qquad\qquad\qquad
    + {1 \over {m-1/3}} \, {\tilde \alpha}_{-(m-1/3)} \,
      e^{i(\tau+\sigma)(m-1/3)}
   \Bigg\}
\end{eqnarray}
 with $(\alpha_{m+1/3})^\dag = \alpha_{-(m+1/3)}$ and
 $({\tilde \alpha}_{m-1/3})^\dag = {\tilde \alpha}_{-(m-1/3)}$.
The quantization results the following algebra:
\begin{equation}
 [ \alpha_{m+1/3}, \alpha_{-(m'+1/3)} ] = (m+1/3) \delta_{m,m'},
\quad
 [ {\tilde \alpha}_{m-1/3}, {\tilde \alpha}_{-(m'-1/3)} ]
  = (m-1/3) \delta_{m,m'}.
\end{equation}
We find no zero mode,
 and the twisted closed string is localized at the singularity.
At the boundary, $\tau=0$,
 the mode expansions can be written as
\begin{eqnarray}
 Z^{(+)}(\sigma,\tau=0)
 &=& \sum_{m=0}^\infty
    {1 \over \sqrt{m+1/3}} \, {\hat x}_{m+1/3} \, e^{i\sigma(m+1/3)}
\nonumber\\
 &+& \sum_{m=1}^\infty
    {1 \over \sqrt{m-1/3}} \, {\hat x}_{m-1/3} \, e^{-i\sigma(m-1/3)},
\\
 Z^{(-)}(\sigma,\tau=0)
 &=& \sum_{m=0}^\infty
    {1 \over \sqrt{m+1/3}} \, {\hat {\bar x}}_{m+1/3} \, e^{-i\sigma(m+1/3)}
\nonumber\\
 &+& \sum_{m=1}^\infty
    {1 \over \sqrt{m-1/3}} \, {\hat {\bar x}}_{m-1/3} \, e^{i\sigma(m-1/3)}.
\end{eqnarray}
We defined boundary coordinate operators as
\begin{eqnarray}
 {\hat x}_{m+1/3} &=& {\tilde a}_{m+1/3} + a_{-(m+1/3)},
\\
 {\hat x}_{m-1/3} &=& a_{m-1/3} + {\tilde a}_{-(m-1/3)},
\\
 {\hat {\bar x}}_{m+1/3} &=& {\tilde a}_{-(m+1/3)} + a_{m+1/3}
                         = ({\hat x}_{m+1/3})^\dag,
\\
 {\hat {\bar x}}_{m-1/3} &=& a_{-(m-1/3)} + {\tilde a}_{m-1/3}
                         = ({\hat x}_{m-1/3})^\dag,
\end{eqnarray}
 where
\begin{eqnarray}
 a_{m\pm1/3}
  &\equiv& i\sqrt{{\alpha'} \over 2} {1 \over \sqrt{m\pm1/3}}
  {\tilde \alpha}_{m\pm1/3},
  \quad {}^{m \ge 0}_{m \ge 1},
\\
 a_{-(m\pm1/3)}
  &\equiv& -i\sqrt{{\alpha'} \over 2} {1 \over \sqrt{m\pm1/3}}
  {\tilde \alpha}_{-(m\pm1/3)},
  \quad {}^{m \ge 0}_{m \ge 1},
\\
 {\tilde a}_{m\pm1/3}
  &\equiv& i\sqrt{{\alpha'} \over 2} {1 \over \sqrt{m\pm1/3}}
  \alpha_{m\pm1/3},
  \quad {}^{m \ge 0}_{m \ge 1},
\\
 {\tilde a}_{-(m\pm1/3)}
  &\equiv& -i\sqrt{{\alpha'} \over 2} {1 \over \sqrt{m\pm1/3}}
  \alpha_{-(m\pm1/3)},
  \quad {}^{m \ge 0}_{m \ge 1}
\end{eqnarray}
 with
\begin{equation}
 [ a_{m\pm1/3}, a_{-(m'\pm1/3)} ] = {{\alpha'} \over 2} \delta_{m,m'},
\quad
 [ {\tilde a}_{m\pm1/3}, {\tilde a}_{-(m'\pm1/3)} ] 
 = {{\alpha'} \over 2} \delta_{m,m'}.
\end{equation}

The contribution to the closed-string boundary state
 corresponding to the open-string Neumann boundary condition
 can be obtained by
 following the prescription reviewed in section \ref{boundary-states}
 using boundary coordinates $x_{m\pm1/3}$ and ${\bar x}_{m\pm1/3}$
 instead of $x^4_m$, $x^5_m$, ${\bar x}^4_m$ and ${\bar x}^5_m$.
The contribution to the closed-string boundary state
 corresponding to the open-string Dirichlet boundary condition
 can be obtained by the dual transformation:
 ${\tilde \alpha}_{\pm(m-1/3)} \rightarrow - {\tilde \alpha}_{\pm(m-1/3)}$.
There is no zero-mode contribution like
 $\delta({\hat x}^4-y^4)\delta({\hat x}^5-y^5)$,
 because there are no operators corresponding to the zero-mode operators
 ${\hat x}^4$ and ${\hat x}^5$ for twisted closed strings.
The exactly the same arguments
 are applied to the world-sheet fields $Z^{(\pm)a=2}$.
The same arguments are also applied to $Z^{(\pm)a=3}$
 with the replacement of $v_{a=1,2}=1/3$ to $v_{a=3}=-2/3$.

The mode expansion of
 twisted world-sheet fermion fields $\psi_\pm^{(\pm)a=1}$,
 we simply write this fields as $\psi_\pm^{(\pm)}$ for a while,
 are given as
\begin{equation}
 \left\{
 \begin{array}{l}
  \psi_+^{(+)}
   = \displaystyle{\sum_{m=-\infty}^\infty}
     \psi^{(+)}_{m+\kappa-1/3} \ e^{-i(\tau+\sigma)(m+\kappa-1/3)},
  \\
  \psi_-^{(+)}
   = \displaystyle{\sum_{m=-\infty}^\infty}
     {\tilde \psi}^{(+)}_{m+\kappa+1/3} \ e^{-i(\tau-\sigma)(m+\kappa+1/3)},
 \end{array}
 \right.
\end{equation}
\begin{equation}
 \left\{
 \begin{array}{l}
  \psi_+^{(-)}
   = \displaystyle{\sum_{m=-\infty}^\infty}
     \psi^{(-)}_{m+\kappa+1/3} \ e^{-i(\tau+\sigma)(m+\kappa+1/3)},
  \\
  \psi_-^{(-)}
   = \displaystyle{\sum_{m=-\infty}^\infty}
     {\tilde \psi}^{(-)}_{m+\kappa-1/3} \ e^{-i(\tau-\sigma)(m+\kappa-1/3)}
 \end{array}
 \right.
\end{equation}
 with $(\psi^{(+)}_{m+\kappa-1/3})^\dag=\psi^{(-)}_{-(m+\kappa-1/3)}$ and
 $({\tilde \psi}^{(+)}_{m+\kappa+1/3})^\dag
   ={\tilde \psi}^{(-)}_{-(m+\kappa+1/3)}$.
There is no zero-mode operator even in Ramond sector.
The quantization results
\begin{equation}
 \{ \psi^{(+)}_r, \psi^{(-)}_s \} = \delta_{r,-s},
\quad
 \{ \psi^{(+)}_s, \psi^{(-)}_r \} = \delta_{s,-r},
\end{equation}
 where, and from now on,
\begin{equation}
 r \in {\bf Z} + \kappa - 1/3,
\quad
 s \in {\bf Z} + \kappa + 1/3.
\end{equation}
The fermionic boundary coordinate operators,
 ${\hat \theta}^{(+)}_r$ and ${\hat \theta}^{(-)}_s$,
 corresponding to the open-string Neumann boundary condition
 are defined as follows
\begin{eqnarray}
 \theta^{(+)}(\sigma;\eta)
  &\equiv& \psi_+^{(+)}(\sigma,\tau=0) - \eta i \psi_-^{(+)}(\sigma,\tau=0),
\\
 \theta^{(-)}(\sigma;\eta)
  &\equiv& \psi_+^{(-)}(\sigma,\tau=0) - \eta i \psi_-^{(-)}(\sigma,\tau=0)
\end{eqnarray}
 with
\begin{equation}
 \theta^{(+)}(\sigma;\eta) = \sum_r {\hat \theta}^{(+)}_r e^{-i\sigma r},
\quad
 \theta^{(-)}(\sigma;\eta) = \sum_s {\hat \theta}^{(-)}_s e^{-i\sigma s}.
\end{equation}
More explicitly,
\begin{equation}
 {\hat \theta}^{(+)}_r
  = \psi^{(+)}_r - \eta i {\tilde \psi}^{(+)}_{-r},
\quad
 {\hat \theta}^{(-)}_s
  = \psi^{(-)}_s - \eta i {\tilde \psi}^{(-)}_{-s}.
\end{equation}
The anti-commuting relations
\begin{equation}
 \{ {\hat \theta}^{(+)}_r, {\hat \theta}^{(+)}_{r'} \} = 0,
\quad
 \{ {\hat \theta}^{(-)}_s, {\hat \theta}^{(-)}_{s'} \} = 0,
\quad
 \{ {\hat \theta}^{(+)}_r, {\hat \theta}^{(-)}_s \} = 0
\end{equation}
 are satisfied independent from the value of $\eta$.
We define
\begin{equation}
 {\hat {\bar \theta}}^{(+)}_r \equiv {\hat \theta}^{(+)}_{-r},
\quad
 {\hat {\bar \theta}}^{(-)}_s \equiv {\hat \theta}^{(-)}_{-s}
\end{equation}
 for $r,s>0$.
The contribution to the closed-string boundary state
 corresponding to the open-string Neumann boundary condition
 can be obtained by using these boundary coordinate operators.
The contribution to the boundary states
 corresponding to the open-string Dirichlet boundary state
 can be obtained by the dual transformation:
 ${\tilde \psi}^{(+)}_s \rightarrow -{\tilde \psi}^{(+)}_s$
 and
 ${\tilde \psi}^{(-)}_r \rightarrow -{\tilde \psi}^{(-)}_r$.

Since there is no zero-mode operator in Ramond sector ($\kappa=0$),
 we should reconstruct the ground state of R-R sector.
For D$p$-brane ($p < 4$) it can be obtained as
\begin{equation}
 \vert B_\psi; \eta \rangle^{(0)}_R
 = (M_{\rm 4D})_{ab} \vert a \rangle_R {\widetilde {\vert b \rangle}_R},
\end{equation}
 where $\vert a \rangle_R$ and ${\widetilde {\vert b \rangle}_R}$
 are four-dimensional spinor states with spinor indices $a$ and $b$,
 and
\begin{equation}
 M_{\rm 4D}
  = C_{\rm 4D} \Gamma^0 \cdots \Gamma^p
    {{1+i\eta\Gamma^5} \over {1+i\eta}}
\end{equation}
 with $C_{\rm 4D} = \Gamma^1 \Gamma^2$ and
 $\Gamma^5 = -i \Gamma^0 \Gamma^1 \Gamma^2 \Gamma^3$.
This state satisfies the Neumann and Dirichlet boundary conditions
 for the zero-mode operators of four-dimensional space-time:
\begin{equation}
 \left( \psi^\mu_0 - i \eta {\tilde \psi}^\mu_0 \right)
  \vert B_\psi; \eta \rangle^{(0)}_R = 0,
  \quad \mbox{for $\mu=0,\cdots,p$},
\end{equation}
\begin{equation}
 \left( \psi^i_0 + i \eta {\tilde \psi}^i_0 \right)
  \vert B_\psi; \eta \rangle^{(0)}_R = 0,
  \quad \mbox{for $i=p+1,\cdots,3$}.
\end{equation}

A special care is required
 for the normalization factor of this D$p$-brane boundary state.
The normalization factor of the D$p$-brane boundary state is determined
 so that the open-closed string duality is satisfied.
Namely, it is determined from the condition that
 the vacuum amplitude which is obtained using D$p$-brane boundary states
 should coincide with the twice of the open-string one-loop vacuum amplitude
 which is calculated using the open-string world-sheet formalism.
Consider the case of
 a stack of D$3$-branes at a ${\bf C}^3/{\bf Z}_3$ singularity.
The normalization factor is not that simple $N_3$, but
\begin{equation}
 \sqrt{
  {1 \over 3}
  \left( {1 \over \sqrt{2\pi\alpha'}} \right)^6
  \left| \prod_{a=1}^3 2 \sin(\pi v_a) \right|
 }
 \cdot
 {\rm tr}\left( \gamma_3^{-1} \right)
 \times N_3
 \equiv N_3^{\rm T} \times {\rm tr} \left( \gamma_3^{-1} \right),
\label{normalization-twisted}
\end{equation}
 where $\gamma_3$ is the matrix of ${\bf Z}_3$ operation on
 open-string Chan-Paton indices.
For the adjoint state the factor ${\rm tr}(\gamma_3^{-1})$
 should be replaced by ${\rm tr}(\gamma_3)$.
The factor $1/3$ in the square root of Eq.(\ref{normalization-twisted})
 comes from the ${\bf Z}_3$ projection operator
 $(1 + {\hat \alpha} + {\hat \alpha}^2)/3$,
 where ${\hat \alpha}$ is the operator
 which generates the transformation of Eq.(\ref{Z3-transformation}).
A dimension-full factor $\left( 1 / \sqrt{2\pi\alpha'} \right)^6$
 comes instead of the momentum integration in six-dimensional space
 perpendicular to the three-dimensional space of D$3$-branes.
The last factor in the square root of Eq.(\ref{normalization-twisted})
 has already been introduced in Ref.\cite{Diaconescu:1999dt}
 with a certain interpretation.

Now,
 we construct the boundary state of the D$3$-branes
 at a supersymmetric ${\bf C}^3/{\bf Z}_3$ singularity
 with the constant background field
 by introducing the boundary action $S_A^\perp$
 of Eq.(\ref{boundary-action-moduli}).
Consider the background of $A \equiv (A_4 + i A_5)/\sqrt{2}$
 with $A = a T$, where $T$ is a Chan-Paton matrix.
Since the matrix $T$ should non-trivially transform by $\gamma_3$
 so that the state corresponding to $A$
 is invariant under the ${\bf Z}_3$ transformation,
 we have ${\rm tr}(T)=0$.
The field $a$ is not the brane moduli field,
 but a matter field in some non-trivial representation of the gauge group.
The boundary action is described by
 dual bosonic boundary coordinates for $Z^{(\pm)a=1}$ as follows.
\begin{eqnarray}
 S_A^\perp
 = &+& \left( e^{i 2 \pi / 3} - 1 \right) A^\dag
   \left\{
    \sum_{m=0}^\infty {{x_D \, {}_{m+1/3}} \over \sqrt{m+1/3}}
    + \sum_{m=1}^\infty {{x_D \, {}_{m-1/3}} \over \sqrt{m-1/3}}
   \right\}
\nonumber\\
   &-& \left( e^{-i 2 \pi / 3} - 1 \right) A
   \left\{
    \sum_{m=0}^\infty {{{\bar x}_D \, {}_{m+1/3}} \over \sqrt{m+1/3}}
    + \sum_{m=1}^\infty {{{\bar x}_D \, {}_{m-1/3}} \over \sqrt{m-1/3}}
   \right\}.
\end{eqnarray}
Although $A$ (and $S_A^\perp$) is the matrix valued,
 we replace $A$ by $a$ for a while,
 and consider the effect of the Chan-Paton matrix afterward.
The boundary state before GSO projection
 is obtained almost the same as Eq.(\ref{boundary-with-A})
 by replacing $S_A$ by the above $S_A^\perp$
 with some special cares described above.
The products in the eigenstates of boundary coordinates
 are modified as
\begin{eqnarray}
 ({\bar x}|x)_\parallel
  &\equiv& {2 \over {\alpha'}}
         \sum_{m=1}^\infty
         \sum_{\mu,\nu=0}^3 \eta_{\mu\nu} {\bar x}^\mu_m x^\nu_m,
\\
 ({\bar x}_D|x_D)_\perp
  &\equiv& {2 \over {\alpha'}}
        \sum_{a=1}^3
        \left\{
         \sum_{m=0}^\infty
          {\bar x}_D{}^a_{m+1/3} x_D{}^a_{m+1/3}
       + \sum_{m=1}^\infty
          {\bar x}_D{}^a_{m-1/3} x_D{}^a_{m-1/3}
        \right\},
\\
 ({\bar \theta}|\theta)_\parallel
  &\equiv& \sum_{r>0}^\infty
         \sum_{\mu,\nu=0}^3 \eta_{\mu\nu} {\bar \theta}^\mu_r \theta^\nu_r,
\\
 ({\bar \theta}_D|\theta_D)_\perp
  &\equiv&
         \sum_{a=1}^3
         \left\{
          \sum_{r>0}^\infty {\bar \theta}_D{}^{(+)a}_r \theta_D{}^{(+)a}_r
        + \sum_{s>0}^\infty {\bar \theta}_D{}^{(-)a}_s \theta_D{}^{(-)a}_s
         \right\},
\end{eqnarray}
 and the same for the others.

The GSO projection of the state can be defined
 as described in section \ref{introduction},
 like Eqs.(\ref{GSO-NS}) and (\ref{GSO-R}),
 but a special care is required.
Since the twisted world-sheet fermion fields
 have non-trivial conformal weights,
 the corresponding vacuum states
 should have non-trivial world-sheet fermion number, or GSO parity.
For NS-NS sector
 the vacuum state has GSO parity $\exp(i\pi(1/3+1/3+7/3))=-1$
 and Eq.(\ref{GSO-NS}) is changed as
\begin{equation}
 {{1+{e^{i \pi F}}} \over 2}{{1+{e^{i \pi {\tilde F}}}} \over 2}
 \vert B_p ; +1 \rangle_{\rm NS}
 = {1 \over 2}
   \left(
    \vert B_p ; +1 \rangle_{\rm NS}
    + \vert B_p ; -1 \rangle_{\rm NS}
   \right)
   \equiv \vert B_p \rangle_{\rm NS}.
\end{equation}
For R-R sector
 the vacuum state has GSO parity $\exp(i\pi(2/3+2/3-4/3))=+1$
 and Eq.(\ref{GSO-R}) is not changed.

Since the boundary action is linear in dual boundary coordinates,
 we can explicitly perform the following part of the functional integrals.
\begin{equation}
 \vert X^{4,5} \rangle =
 \int {\cal D}{\bar x}_D^{+1/3}{\cal D}x_D^{+1/3}
      {\cal D}{\bar x}_D^{-1/3}{\cal D}x_D^{-1/3}
 \ e^{iS_A^\perp} \
 e^{-{1 \over 2}({\bar x}_D \vert x_D)
    + (a^\dag \vert {\tilde a}^\dag)
    - (a^\dag \vert x_D)
    + ({\bar x}_D \vert {\tilde a}^\dag)}
 \vert 0 \rangle,
\end{equation}
 where
\begin{equation}
 {\cal D}{\bar x}_D^{+1/3}{\cal D}x_D^{+1/3}
 {\cal D}{\bar x}_D^{-1/3}{\cal D}x_D^{-1/3}
  \equiv \prod_{m=0}^\infty d{\bar x}_D \, {}_{m+1/3}
                          \ dx_D \, {}_{m+1/3}
         \prod_{m=1}^\infty d{\bar x}_D \, {}_{m-1/3}
                          \ dx_D \, {}_{m-1/3}.
\end{equation}
The result is
\begin{equation}
 \vert X^{4,5} \rangle =
 e^{- ( a^\dag \vert {\tilde a}^\dag)
    + 2 ( a^\dag \vert \Phi )
    + 2 ( \Phi^\dag \vert {\tilde a}^\dag )
    - 2 ( \Phi^\dag \vert \Phi )}
 \vert 0 \rangle
\label{boundary-state-Phi}
\end{equation}
 with
\begin{equation}
 \Phi_{m \pm 1/3}
  \equiv i {{\alpha'} \over 2}
         {{e^{-i 2 \pi / 3} - 1} \over \sqrt{m \pm 1/3}} \ g \phi,
\qquad
 \Phi_{m \pm 1/3}^\dag
  \equiv i {{\alpha'} \over 2}
         {{e^{+i 2 \pi / 3} - 1} \over \sqrt{m \pm 1/3}} \ g \phi^\dag,
\end{equation}
 where we take the canonical normalization of the scalar field as
 $\phi \equiv a/g$ with $1/g^2=T_3^{\rm DBI} (2 \pi \alpha')^2/g_s$
 assuming the normalization of ${\rm tr} (T^\dag T) = 1$.
This is the only modified part
 in the twisted closed-string D$3$-brane boundary state
 at a ${\bf C}^3/{\bf Z}_3$ singularity
 by this constant open-string background.
The boundary state with this open-string background can be described as
\begin{equation}
 \vert B_3; \phi \rangle =
  N_3^{\rm T}
   e^{(a^\dag \vert {\tilde a}^\dag)_\parallel}
   e^{- ( a^\dag \vert {\tilde a}^\dag)
      + 2 ( a^\dag \vert \Phi )
      + 2 ( \Phi^\dag \vert {\tilde a}^\dag )
      - 2 ( \Phi^\dag \vert \Phi )}
   e^{-(a^\dag \vert {\tilde a}^\dag)_{\perp'}}
  \vert 0 \rangle
  \vert B_3^\psi \rangle
  \vert B_{\rm gh} \rangle
  \vert B_{\rm sgh} \rangle,
\end{equation}
 where $\vert B_3^\psi \rangle$ is the world-sheet fermion contribution
 and $(a^\dag \vert {\tilde a}^\dag)_{\perp'}$ does not include
 the contribution from the space component of $a=1$.
The scalar background field appears in a more complicated way
 than in Eq.(\ref{boundary-state-bosonic-with-scalar}).

The ``multipoint function'' can be obtained from
 $\langle B_3; \phi \vert D \vert B_3; \phi \rangle$.
For the ``two-point function'', we have a simple formula
\begin{eqnarray}
 A_2 &=& \left\{
        {\rm tr}(\gamma_3 T^\dag T){\rm tr}(\gamma_3^{-1})
        + {\rm tr}(\gamma_3){\rm tr}(T^\dag T \gamma_3^{-1})
       \right\}
\nonumber\\
 && \quad
       \times
       \left\{ - 2 ( \Phi^\dag \vert \Phi ) \right\}
       \langle B_3; \phi=0 \vert
        D
       \vert B_3; \phi=0 \rangle,
\end{eqnarray}
 where we included the Chan-Paton factor
 which appears in the open-string one-loop calculation.
The linear terms of $\Phi$ in the exponent of Eq.(\ref{boundary-state-Phi})
 do not contribute,
 because in open string picture the amplitude with only one vertex operator
 on one boundary vanishes due to ${\rm tr}(T \gamma_3)=0$
 (or gauge invariance).
We see that
 the amplitude is proportional to
 the open-string one-loop vacuum amplitude with a twist,
 which is dual of the amplitude of a tree-level propagation
 of the twisted closed-string.
The factor $- 2 ( \Phi^\dag \vert \Phi )$ is a divergent quantity
 that requires regularization through the analytic continuation.
\begin{eqnarray}
 - 2 ( \Phi^\dag \vert \Phi )
 &=& 2 \cdot {{\alpha'} \over 2} \left| e^{i2\pi/3}-1 \right|^2
     g^2 \phi^\dag \phi
     \left\{
      \sum_{m=0}^\infty {1 \over {m+1/3}}
      + \sum_{m=1}^\infty {1 \over {m-1/3}}
     \right\}
\nonumber\\
 &=& 3 \alpha' g^2 \phi^\dag \phi
     \left( 2 \gamma + 3 \ln 3 \right).
\end{eqnarray}
We used the polygamma function defined as
\begin{equation}
 \psi_n(z) = (-1)^{n+1} n! \sum_{k=0}^\infty {1 \over {(z+k)^{n+1}}},
\end{equation}
 and
\begin{eqnarray}
 \sum_{m=0}^\infty {1 \over {m+1/3}}
  &=& - \psi_0(1/3) = \gamma + {{\pi\sqrt{3}} \over 6} + {{3 \ln 3} \over 2},
\\
 \sum_{m=1}^\infty {1 \over {m-1/3}}
  &=& - \psi_0(2/3) = \gamma - {{\pi\sqrt{3}} \over 6} + {{3 \ln 3} \over 2}.
\end{eqnarray}
The mass of the scalar field in D$3$-brane world-volume is obtained as
\begin{equation}
 m_\phi^2 = - 3 g^2 \left( 2 \gamma + 3 \ln 3 \right)
            \left\{
             {\rm tr}(\gamma_3 T^\dag T){\rm tr}(\gamma_3^{-1})
             + {\rm tr}(\gamma_3){\rm tr}(T^\dag T \gamma_3^{-1})
            \right\}
            \times \alpha' A_{\rm vac}^{\rm T}/V_4
\end{equation}
 with
\begin{eqnarray}
 A_{\rm vac}^{\rm T}
  &=& 
   {1 \over 3}
   \left(
    {1 \over \sqrt{2 \pi \alpha'}}
   \right)^6
   \left| \prod_{a=1}^3 2 \sin(\pi v_a) \right|
   \cdot
   \left( 2 \pi^2 {{\alpha'} \over {4\pi}} N_3^2 \right) \cdot V_4
   \int_0^\infty ds {1 \over {(\eta(is))^2}}
\nonumber\\
  && 
   \times 
   {1 \over 2}
   \left\{
     {{\theta\left[\begin{array}{c} 0 \\ 0 \end{array}\right]} \over \eta(is)}
     \left(
      {{\theta\left[\begin{array}{c} 1/3 \\ 0 \end{array}\right]}
       \over
       {\theta\left[\begin{array}{c} 1/6 \\ 1/2 \end{array}\right]
        /e^{2 \pi i \cdot {1 \over 6} \cdot {1 \over 2}}}}
     \right)^3
   + {{\theta\left[\begin{array}{c} 0 \\ 1/2 \end{array}\right]} \over \eta(is)}
     \left(
      {{\theta\left[\begin{array}{c} 1/3 \\ 1/2 \end{array}\right]
        /e^{2 \pi i \cdot {1 \over 3} \cdot {1 \over 2}}}
       \over
       {\theta\left[\begin{array}{c} 1/6 \\ 1/2 \end{array}\right]
        /e^{2 \pi i \cdot {1 \over 6} \cdot {1 \over 2}}}}
     \right)^3
   \right.
\nonumber\\
  && \qquad\qquad\qquad
   \left.
   - {{\theta\left[\begin{array}{c} 1/2 \\ 0 \end{array}\right]} \over \eta(is)}
     \left(
      {{\theta\left[\begin{array}{c} 1/6 \\ 0 \end{array}\right]}
       \over
       {\theta\left[\begin{array}{c} 1/6 \\ 1/2 \end{array}\right]
        /e^{2 \pi i \cdot {1 \over 6} \cdot {1 \over 2}}}}
     \right)^3
   \right\},
\label{B3-B3-vac}
\end{eqnarray}
 where the generalized theta function is defined as \cite{Angelantonj:2002ct}
\begin{eqnarray}
 \theta\left[\begin{array}{c} \alpha \\ \beta \end{array}\right](z|\tau)
 &\equiv&
 e^{2 \pi i \alpha (z + \beta)}
 q^{\alpha^2/2} \prod_{n=1}^\infty \left( 1 - q^n \right)
\nonumber\\
 &\times&
 \prod_{m=1}^\infty
  \left( 1 + q^{m+\alpha-1/2} e^{2 \pi i (z+\beta)} \right)
  \left( 1 + q^{m-\alpha-1/2} e^{-2 \pi i (z+\beta)} \right)
\end{eqnarray}
 with $q = \exp(2 \pi i \tau)$,
 and we used an abbreviation
\begin{equation}
  \theta\left[\begin{array}{c} \alpha \\ \beta \end{array}\right]
  =
  \theta\left[\begin{array}{c} \alpha \\ \beta \end{array}\right](0|is).
\end{equation}
This vacuum amplitude and the mass
 vanish due to the supersymmetry through the identity
\begin{equation}
 \theta\left[\begin{array}{c} 0 \\ 0 \end{array}\right]
  \left( \theta\left[\begin{array}{c} 1/3 \\ 0 \end{array}\right] \right)^3
 -
 \theta\left[\begin{array}{c} 0 \\ 1/2 \end{array}\right]
  \left( \theta\left[\begin{array}{c} 1/3 \\ 1/2 \end{array}\right] \right)^3
 -
 \theta\left[\begin{array}{c} 1/2 \\ 0 \end{array}\right]
  \left( \theta\left[\begin{array}{c} 1/6 \\ 0 \end{array}\right] \right)^3
 = 0.
\end{equation}
This identity is obtained from more general identity \cite{Dabholkar:1994ai}
\begin{eqnarray}
 2 \prod_{i=1}^4
  \theta\left[\begin{array}{c} 1/2 \\ 1/2 \end{array}\right](x_i|\tau)
 &=&
  \prod_{i=1}^4
  \theta\left[\begin{array}{c} 0 \\ 0 \end{array}\right](y_i|\tau)
 -
 \prod_{i=1}^4
  \theta\left[\begin{array}{c} 0 \\ 1/2 \end{array}\right](y_i|\tau)
\nonumber\\
 &-&
 \prod_{i=1}^4
  \theta\left[\begin{array}{c} 1/2 \\ 0 \end{array}\right](y_i|\tau)
 +
 \prod_{i=1}^4
  \theta\left[\begin{array}{c} 1/2 \\ 1/2 \end{array}\right](y_i|\tau),
\end{eqnarray}
 where
\begin{equation}
 \left\{
 \begin{array}{l}
  y_1 = (x_1 + x_2 + x_3 + x_4)/2 \\
  y_2 = (x_1 - x_2 - x_3 + x_4)/2 \\
  y_3 = (x_1 + x_2 - x_3 - x_4)/2 \\
  y_4 = (x_1 - x_2 + x_3 - x_4)/2
 \end{array}
 \right..
\end{equation}

Once the supersymmetry is broken,
 the modified spectrum gives finite values of the mass squared
 with sign determined by the signs of the vacuum energy and Chan-Paton factor.
The scale of the mass
 is determined only by the string scale $1/\sqrt{\alpha'}$
 in case of that the supersymmetry is broken by construction, for example,
 by non-supersymmetric singularities \cite{Aldazabal:2000sa},
 non-supersymmetric brane configurations \cite{Antoniadis:1999xk}
 and so on.
In case of that the supersymmetry is spontaneously broken
 by some dynamics at lower energy,
 the scale of the mass is determined by the string scale
 and supersymmetry breaking scale
 which may be introduced through the dimensional transmutation,
 for example.

\section{One-loop masses with brane SUSY breaking}
\label{brane-susy-br}

In this section
 we examine the non-trivial (non-vanishing) one-loop scalar mass
 in a consistent concrete non-tachyonic model
 with brane supersymmetry breaking
 \cite{Sugimoto:1999tx,Antoniadis:1999xk,Angelantonj:1999jh,
 Aldazabal:1999jr,Angelantonj:1999ms}.
Consider the same ${\bf C}^3/{\bf Z}_3$ orbifold singularity
 in the previous section,
 and put four D$3$-branes and three anti-D$7$-brane
 at the singularity.
We take the ${\bf Z}_3$ operation matrices to these D-branes as
\begin{eqnarray}
 \gamma_3
  &=& {\rm diag}({\bf 1}_2, \alpha {\bf 1}_1, \alpha^2 {\bf 1}_1),
\\
 \gamma_{{\bar 7}_3} &=& {\bf 1}_3,
\end{eqnarray}
 where $\alpha = \exp ( i 2 \pi / 3)$,
 ${\bf 1}_a$ is $a \times a$ unit matrix.
The subscript ${\bar 7}_3$ means
 the anti-D$7$-brane which does not occupy
 the space dimensions of third complex coordinate.
The twisted R-R tadpole cancellation condition
\begin{equation}
 3 {\rm Tr} (\gamma_3) - {\rm Tr} (\gamma_{{\bar 7}_3}) = 0
\end{equation}
 is satisfied
 and the gauge symmetry, U$(2) \times$U$(1)_1 \times$U$(1)_2 \times$U$(3)$,
 should be anomaly free.
Though each of two U$(1)$ gauge symmetries,
 U$(1)_1$ and U$(1)_2$, on D$3$-brane is ``anomalous U$(1)$'' gauge symmetry
 whose anomaly is cancelled out by generalized Green-Schwarz mechanism
 \cite{Sagnotti:1992qw},
 the diagonal U$(1)$ gauge symmetry of them is
 ``non-anomalous U$(1)$'' gauge symmetry.
Since the gauge boson of anomalous U$(1)$ gauge symmetry becomes massive,
 we consider only the non-anomalous diagonal U$(1)$ gauge symmetry
 in the following.

The massless spectrum on four-dimensional D$3$-brane world-volume
 is the following.
There are
 ${\cal N}=1$ supersymmetry gauge multiplets of U$(2) \times$U$(1)$
 and chiral multiplets with the following
 U$(2) \times$U$(3) \times$U$(1)$ quantum numbers:
\begin{eqnarray}
 (\Phi_1^a, \Psi_1^a): &\quad& (2^*, 1)_{+1},
\\
 (\Phi_2^a, \Psi_2^a): &\quad& (1, 1)_{0},
\\
 (\Phi_3^a, \Psi_3^a): &\quad& (2, 1)_{-1},
\end{eqnarray}
 where $a=1,2,3$.
The D$3$-D${\bar 7}_3$ and D${\bar 7}_3$-D$3$ open strings
 give non supersymmetric spectrum:
\begin{eqnarray}
 \phi_1: &\quad& (2, 3^*)_{0},
\\
 \phi_2: &\quad& (2^*, 3)_{0},
\\
 \psi_1: &\quad& (1, 3)_{-1},
\\
 \psi_2: &\quad& (1, 3^*)_{+1},
\end{eqnarray} 
 where $\phi_{1,2}$ are massless complex scalar fields
 and $\psi_{1,2}$ are Weyl fermion fields.

We estimate one-loop mass of the complex scalar field $\Phi_2^3$
 which we simply call $\phi$ in this section.
Note that this scalar field
 is singlet under all the non-anomalous gauge symmetries,
 and the anomalous U$(1)$ gauge interaction should give
 supersymmetric, namely zero, contribution to the one-loop mass.
Non-trivial contributions
 should come from the following possible non-supersymmetric interactions:
\begin{equation}
 \phi (\psi_1 \psi_2),
 \qquad
 (\phi^\dag \phi) (\phi_1^\dag \phi_1),
 \qquad
 (\phi^\dag \phi) (\phi_2^\dag \phi_2).
 \qquad
\end{equation}
The fermion loop due to the first interaction
 should give negative contributions to the mass squared,
 and the boson loop due to the second and third interactions
 should give positive contributions.
Since the calculation of the coupling constants of these interactions
 is beyond the scope of this article,
 we do not make an observation on the sign of mass squared
 from low energy field theoretical point of view.

The calculation in String Theory is almost the same in the previous section.
The result is
\begin{equation}
 m_\phi^2
  = 9 g^2 (2 \gamma + 3 \ln 3) \alpha'
    \left[
     3 \langle B_{{\bar 7}_3} \vert D \vert B_3 \rangle
     + \langle B_3 \vert D \vert B_3 \rangle
    \right]/V_4,
\label{mass-brane-susy-br}
\end{equation}
 where the vacuum amplitude $\langle B_3 \vert D \vert B_3 \rangle$
 has already been given in Eq.(\ref{B3-B3-vac}) and
 the vacuum amplitude $\langle B_{{\bar 7}_3} \vert D \vert B_3 \rangle$
 is given as
\begin{eqnarray}
 {\bar A}_{\rm vac}^{\rm T}
  &=& 
   {1 \over 3}
   \left(
    {1 \over \sqrt{2 \pi \alpha'}}
   \right)^6
   \left| 2 \sin(\pi v_3) \right|
   \cdot
   \left( 2 \pi^2 {{\alpha'} \over {4\pi}} N_3^2 \right) \cdot V_4
   \int_0^\infty ds {1 \over {(\eta(is))^2}}
\nonumber\\
  && 
   \times 
   {1 \over 2}
   \left\{
     {{\theta\left[\begin{array}{c} 0 \\ 0 \end{array}\right]} \over \eta(is)}
     \left(
      {{\theta\left[\begin{array}{c} 1/3 \\ 1/2 \end{array}\right]
        /e^{2 \pi i \cdot {1 \over 3} \cdot {1 \over 2}}}}
       \over
       {\theta\left[\begin{array}{c} 1/6 \\ 0 \end{array}\right]}
     \right)^2
       {{\theta\left[\begin{array}{c} 1/3 \\ 0 \end{array}\right]}
       \over
       {\theta\left[\begin{array}{c} 1/6 \\ 1/2 \end{array}\right]
        /e^{2 \pi i \cdot {1 \over 6} \cdot {1 \over 2}}}}
   \right.
\nonumber\\
  && \qquad\qquad
   + {{\theta\left[\begin{array}{c} 0 \\ 1/2 \end{array}\right]} \over \eta(is)}
     \left(
      {{\theta\left[\begin{array}{c} 1/3 \\ 0 \end{array}\right]}
       \over
       {\theta\left[\begin{array}{c} 1/6 \\ 0 \end{array}\right]}}
     \right)^2
      {{\theta\left[\begin{array}{c} 1/3 \\ 1/2 \end{array}\right]}
        /e^{2 \pi i \cdot {1 \over 3} \cdot {1 \over 2}}
       \over
       {\theta\left[\begin{array}{c} 1/6 \\ 1/2 \end{array}\right]
        /e^{2 \pi i \cdot {1 \over 6} \cdot {1 \over 2}}}}
\nonumber\\
  && \qquad\qquad
   \left.
   + {{\theta\left[\begin{array}{c} 1/2 \\ 0 \end{array}\right]} \over \eta(is)}
     \left(
      {{\theta\left[\begin{array}{c} 1/6 \\ 1/2 \end{array}\right]
        /e^{2 \pi i \cdot {1 \over 6} \cdot {1 \over 2}}}
       \over
       {\theta\left[\begin{array}{c} 1/6 \\ 0 \end{array}\right]}}
     \right)^2
      {{\theta\left[\begin{array}{c} 1/6 \\ 0 \end{array}\right]}
       \over
       {\theta\left[\begin{array}{c} 1/6 \\ 1/2 \end{array}\right]
        /e^{2 \pi i \cdot {1 \over 6} \cdot {1 \over 2}}}}
   \right\}.
\label{B3-B7-vac}
\end{eqnarray}
The third terms in curly brackets
 of Eqs.(\ref{B3-B3-vac}) and (\ref{B3-B7-vac})
 represent R-R closed string exchanges,
 and the divergent massless tadpole contribution is cancelled out.
Note that the sign of the term in Eq.(\ref{B3-B7-vac})
 is opposite to the case of usual D$7$-branes.
For the aim of the order estimate,
 we expand the integrant for large $s$
 and take only the leading term.
The result is
\begin{equation}
 m_\phi^2|_{\rm R-R}
  = 9 g^2 (2 \gamma + 3 \ln 3) \alpha'
    \cdot (N_3^T)^2 \cdot 2\pi^2 {{\alpha'} \over {4\pi}}
    \cdot \left( - 8 \cdot {3 \over {2 \pi}} \right).
\label{mass-brane-susy-br-RR}
\end{equation}
 
The first and second terms in the curly brackets
 of Eqs.(\ref{B3-B3-vac}) and (\ref{B3-B7-vac})
 represent NS-NS closed string exchanges.
The divergence
 due to the massless states with tadpole couplings to D-branes
 is not cancelled out.
\begin{equation}
 \left[
  3 \langle B_{{\bar 7}_3} \vert D \vert B_3 \rangle
  + \langle B_3 \vert D \vert B_3 \rangle
 \right]_{\rm NS-NS}
 = (N_3^T)^2 \cdot 2\pi^2 {{\alpha'} \over {4\pi}} V_4
   \cdot 2 \int_0^\infty ds
   + \mbox{finite}.
 \end{equation}
The massless fields from NS-NS twisted sector
 with tadpole couplings, are identified as real scalar fields
 in Refs.\cite{Merlatti:2000ne,Bertolini:2001gg},
 and the low energy effective action of the leading order is given.
The tadpole resummation procedure,
 which has been proposed in Ref.\cite{Kitazawa:2008hv},
 is possible in case that
 the higher order contact interactions between the scalar fields and D-branes
 are known.
In the present case,
 as far as the author knows,
 the higher order interactions beyond the tadpole couplings are not known
 except for a suggestion of two-point contact interaction
 in Ref.\cite{Marotta:2002yg}.
Although we should stop here to keep the strictness of calculations,
 it may be useful to sketch
 how the tadpole resummation could actuary be carried out
 assuming a contact two-point interaction.
Note that the following calculation
 is not the rigorous one but a demonstration.

From Ref.\cite{Marotta:2002yg} we expect that
 the coupling constant of the two-point contact interaction
 is proportional to the D-brane tension.
Here, we assume that the coupling is given by
\begin{equation}
 T_3^T = 2 N_3^T / \kappa_{\rm 4D},
\end{equation}
 where $1/\kappa_{4D}$
 is the reduced Planck mass in four dimensional world-volume of D$3$-brane.
The tree-level effect of the two-point interaction
 can be included by a kind of Dyson resummation.
Namely,
\begin{eqnarray}
 \langle B_3 \vert D \vert B_3 \rangle
 &\rightarrow&
 \langle B_3 \vert D_M \vert B_3 \rangle
\nonumber\\
 &\equiv& \langle B_3 \vert D \vert B_3 \rangle
        + \langle B_3 \vert D {\hat M} D \vert B_3 \rangle
        + \langle B_3 \vert D {\hat M} D {\hat M} D \vert B_3 \rangle
        + \cdots,
\\
 \langle B_{{\bar 7}_3} \vert D \vert B_3 \rangle
 &\rightarrow&
 \langle B_{{\bar 7}_3} \vert D_M \vert B_3 \rangle
\nonumber\\
 &\equiv& \langle B_{{\bar 7}_3} \vert D \vert B_3 \rangle
        + \langle B_{{\bar 7}_3} \vert D {\hat M} D \vert B_3 \rangle
        + \langle B_{{\bar 7}_3} \vert D {\hat M} D {\hat M} D \vert B_3 \rangle
        + \cdots,
\end{eqnarray}
 where operator ${\hat M}$ is defined as
\begin{equation}
 {\hat M} \equiv \int d^{10}x \delta^6(x)
  \vert {\tilde B}_3 \rangle (- T_3^T) \langle {\tilde B}_3 \vert
\end{equation}
 which describes one insertion of the two-point interaction
 (or one bounce of the closed string on D3-brane).
See Ref.\cite{Kitazawa:2008hv}
 for the details of the definition of this operator.
The explicit calculations give
\begin{eqnarray}
 \langle B_3 \vert D_M \vert B_3 \rangle
 &=& V_4 (N_3^T)^2
  {{\Delta} \over {1 + T_3^T (N_3^T/T_3^T)^2 \Delta}}
  \rightarrow V_4 T_3^T,
\\
 \langle B_{{\bar 7}_3} \vert D_M \vert B_3 \rangle
 &=& V_4 N_{{\bar 7}_3}^T N_3^T
  {{\Delta'} \over {1 + T_3^T (N_3^T/T_3^T)^2 \Delta}}
  \rightarrow V_4 T_3^T {{N_{{\bar 7}_3}^T} \over {N_3^T}},
\end{eqnarray}
 where
\begin{eqnarray}
 \Delta
  &\equiv& \langle B_3 \vert D \vert B_3 \rangle
           / V_4 (N_3^T)^2,
\\
 \Delta'
  &\equiv& \langle B_{{\bar 7}_3} \vert D \vert B_3 \rangle
           / V_4 N_{{\bar 7}_3}^T N_3^T
\end{eqnarray}
 include the same order of divergence with ultraviolet cutoff on $s$,
 and
\begin{equation}
 N_{{\bar 7}_3}^T
 = \sqrt{
   {1 \over 3}
   \left( {1 \over \sqrt{2\pi\alpha'}} \right)^6 (4 \pi^2 \alpha')^4
   \left|
    {{2 \sin (\pi v_3)} \over
     {2 \sin (\pi v_1) \cdot 2 \sin (\pi v_2)}}
   \right|
   } \times N_7.
\end{equation}
This NS-NS tadpole resummation gives
\begin{equation}
 m_\phi^2|_{\rm NS-NS}
  = 9 g^2 (2 \gamma + 3 \ln 3) \alpha'
    \left[
     3 {{N_{{\bar 7}_3}^T} \over {N_3^T}} + 1
    \right] T_3^T
  = 9 g^2 (2 \gamma + 3 \ln 3) \alpha' \cdot 2 T_3^T.
\end{equation}
In total
\begin{equation}
 m_\phi^2
  = 9 g^2 (2 \gamma + 3 \ln 3) \alpha'
    \left[
     {{4N_3^T} \over {\kappa_{\rm 4D}}}
     - (N_3^T)^2
       \cdot 2 \pi^2 {{\alpha'} \over {4\pi}}
       \cdot {{8 \cdot 3} \over {2\pi}}
    \right].
\end{equation}
For quantitative arguments,
 we need to know more precise information
 on couplings of contact interactions
 between massless twisted NS-NS fields and D-branes.

\section{Conclusions}
\label{conclusions}

The massless scalar states from the open strings,
 which are constrained to end on one stack of D-branes,
 are categorized in two types:
 one related with Wilson line degrees of freedom
 and the other related with brane moduli.
The open-string one-loop corrections to the masses of such scalar states
 can be calculated rather simply in the closed-string picture
 using the boundary state formalism.
In case with broken supersymmetry,
 one-loop corrections give non-zero masses
 which have strong correlations with the values of
 the corresponding open-string one-loop vacuum amplitudes,
 or vacuum energies.
The sign of mass squared
 is essentially determined by the sign of the vacuum amplitude.
This may give a hint to construct explicit models
 with radiative gauge symmetry breaking in String Theory.
 
It should be noted that
 the method in this article can not be directly applied
 to the massless scalar states from open strings
 whose two edges end on different stacks of D-branes.
This kind of open strings are important
 in intersecting D-brane models, for example.
Though,
 the method can be applied to investigate the fate of
 the massless adjoint scalar fields,
 further developments are required to discuss the fate of
 other massless scalar fields in intersecting D-brane models. 

In the string models with broken supersymmetry,
 like the model in section \ref{brane-susy-br},
 there is a problem so called NS-NS tadpole problem
 \cite{Fischler:1986ci,Fischler:1986tb,Das:1986dy}.
The existence of uncanceled NS-NS tadpole couplings to D-branes,
 which happens in many string models without supersymmetry,
 means that the background metric and fields (B-field and dilaton field)
 are not the solution of String Theory.
The actual difficulty in calculations
 is the appearance of the divergences in open-string one-loop corrections.
Tadpole resummation \cite{Dudas:2004nd}
 may be applied to avoid NS-NS tadpole problem 
 with the method developed in Ref.\cite{Kitazawa:2008hv}
 using the boundary state formalism.
A sketch of the application is given in section \ref{brane-susy-br}.

It would be interesting to use the method in this article
 as a guide to construct explicit models
 with radiative gauge symmetry breaking.
It would be much more interesting
 to construct pseudo-realistic TeV-scale string models
 with the radiative electroweak symmetry breaking,
 and to discuss their general phenomenology
 at future colliders and other experiments and observations.

\section*{Acknowledgments}

This research was supported in part by
 the Grant-in-Aid for Scientific Research No.~19540303 from
 the Ministry of Education, Culture, Sports, Science and Technology of Japan.

\end{document}